\documentstyle[aps,preprint,epsf]{revtex}

\begin{document}
\draft

\title{Structural Order in Glassy Water}

\author{Nicolas Giovambattista$^1$, Pablo G. Debenedetti$^1$, \\
Francesco Sciortino$^2$, and H. Eugene Stanley$^3$}

\address{
$^1$ Department of Chemical Engineering,\\
Princeton University, Princeton, NJ 08544-5263 USA\\
$^2$ Dipartimento di Fisica and INFM Udr and 
SOFT: Complex Dynamics in Structured Systems\\
Universita' di Roma ``La Sapienza'' -- 
Piazzale Aldo Moro 2, I-00185, Roma, Italy\\
$^3$Center for Polymer Studies and Department of Physics\\
Boston University, Boston, MA 02215 USA \\
}

\date{22 February 2005}

\maketitle 

\begin{abstract}
We  investigate structural order in glassy water by performing
classical molecular dynamics simulations using the extended simple point charge
 (SPC/E) model of water.
  We perform isochoric cooling
 simulations across the glass transition
 temperature at different cooling rates and densities.
We quantify structural order by orientational and translational order metrics.
Upon cooling the liquid into the glassy state,
 both the orientational order parameter $Q$ and
translational order parameter $\tau$ increase. At $T=0$~K, the glasses fall
on a line in the $Q$-$\tau$ plane or {\it order map}.
 The position of this line depends only on density and coincides with the
 location in the order map of the inherent structures (IS) sampled upon cooling.
We evaluate the energy of the IS, $e_{IS}(T)$, and 
find that both order parameters for the IS are proportional
to $e_{IS}$.
We also study the structural order during the transformation of
low-density amorphous ice (LDA) to high-density amorphous ice (HDA) upon
isothermal compression and are able to identify distinct regions in the order
map corresponding to these glasses. Comparison of the 
order parameters for LDA and HDA with those obtained upon isochoric
cooling indicates major structural differences between glasses obtained by
cooling and glasses obtained by compression. These structural differences are
only weakly reflected in the pair correlation function. We also characterize
the evolution of structural order upon isobaric annealing, leading at high
pressure to very-high density amorphous ice (VHDA).
\end{abstract}



\section{Introduction}

Crystalline solids are characterized by a periodic structure and, therefore,
by the presence of long range-oder. A well-defined theoretical and
experimental framework is available for characterizing the structure of crystals
\cite{kittel}. On the other hand, numerous materials found 
in nature and widely utilized in technology exist in an amorphous state
 (e.g., glasses and liquids) \cite{zallen}.
Characterizing the structure of amorphous materials is a challenging
task because such systems organize over short distances and show no
long-range order. Moreover, the local structure in such materials is not
unique and it can change throughout the sample.
Therefore, new tools  are required to characterize the structure of
amorphous materials. One promising approach is to develop order
metrics (order parameters) that quantify certain statistical properties of the
structure \cite{torquato}.  Recent work in this direction has been performed 
on the hard-sphere \cite{hards1,hards2} and Lennard-Jones (LJ)
\cite{LennardJ} particle systems.  Structural order was characterized 
with two metrics:
 a translational order parameter $\tau$ quantifying the tendency of particle pairs
 to adopt preferential separations, and a bond-orientational order
 parameter $Q$ quantifying correlations between the bond angles defined by a 
given particle and its nearest neighbors.
The novelty of this approach resides not in the order metrics themselves,
which have been used before (e.g., \cite{nelson}), but in the 
idea of an {\it order
  map}, in which different states are mapped onto a plane whose axes
represent the order metrics \cite{hards1,hards2}.
The order map for the Lennard-Jones and hard-sphere systems consists of a
line, indicating that the two order metrics are not independent
 \cite{hards2,LennardJ}.

Extension of these ideas to liquid water has proven to be most useful.
In Ref.~\cite{jeff}, the structural order of
cold liquid water was investigated and found to be closely related 
 to the well-known dynamic
\cite{6to9,francislong,AntonioNature} and thermodynamic \cite{pabloBook} 
anomalies of this substance.  It was found that there exists a
 region in the phase diagram bounded by loci of maximum orientational order
  and minimum translational order within which
the dynamic and thermodynamic anomalies occur.
Water's order map was found to be more complex than for simpler,
spherically-symmetric systems. State points define a {\it two-dimensional
  region}  
in the order map, implying that orientational and translational order are
in general independent. Remarkably, however, it is found that the region where water
exhibits dynamic or thermodynamic anomalies defines a {\it line} in the order
map, which implies that when water behaves anomalously, its translational and
orientational order metrics are no longer independent, but are instead
strictly correlated. This property was then used to quantify threshold values
of the order metrics needed in order for various anomalies to occur \cite{jeff}.
Similar (though not identical) behavior was found to occur in another
tetrahedral liquid, silica \cite{qt-silica}.

Notable among water's peculiarities is the fact that it exhibits polyamorphism
 \cite{PabloReview,AngellReview,geneNature}, i.e., the presence
 of more than one glassy form. Uniaxial
compression of ice-$I_h$\cite{mishima84} or
ice-$I_c$\cite{floriano,joharyIc} at $T=77$~K to pressures 
$P \gtrsim 1$~GPa produces a disordered high-density material named
 high-density amorphous ice (HDA). If HDA is recovered at $T=77$~K and
 $P=1$~bar and then heated isobarically, it transforms irreversibly at $T
 \approx 125$~K to a disordered low-density material named 
 low-density amorphous ice (LDA) \cite{mishima84,mishimaNature85}.  
Moreover, LDA transforms to HDA when compressed at $T=77$~K to
$P=0.6$~GPa\cite{mishimaNature85,mishimaVisual}.
The LDA-HDA transformation is reversible above $T\approx
130$~K\cite{mishimaJCP}. Furthermore, because it is
very sharp and shows hysteresis, it has been suggested that the LDA-HDA
transformation  is a
first-order transition \cite{mishimaJCP,mishimaLL,newKlotz}. 
The LDA-HDA transition plays a central role in one of the
thermodynamically-consistent scenarios that have been proposed to explain the
experimentally observed properties of supercooled water \cite{geneNature}.

There are several other routes to producing amorphous ice
\cite{PabloReview,AngellReview,geneNature} in addition to  those mentioned above.
Because glasses are non-equilibrium materials, their properties are
history-dependent, and the specific procedure followed in the glass
 preparation may lead to different
 amorphous forms. It is not clear  whether all the different
amorphous ices can be classified into two {\it families} 
\cite{mishima96,ourVHDAlong,johary2004}, one corresponding to LDA and
 the other to HDA. One complication standing in the way of such a satisfyingly
 simple view is the experimental observation that HDA, when annealed or
 heated at high pressure becomes appreciably denser. The resulting glass is
 called very-high density amorphous ice (VHDA) or relaxed HDA (RHDA)
 \cite{science}. Whether VHDA/RHDA is simply yet another form of glassy water or
 whether, as suggested by simulations \cite{parrinello,ourVHDA}, it is the more
 stable form to which HDA relaxes irreversibly is a question currently
 under investigation.

It should be evident from the preceding discussion that a precise
characterization of structural order in water glasses is of considerably
interest. The purpose of the present work is to provide such a 
 characterization. In particular, 
 (i) we use the order metric approach  introduced in \cite{jeff} for
 equilibrium liquid water  
to quantify structural changes that occur when liquid
water is cooled isochorically to the glassy state; (ii) we quantify the
changes in structural order that accompany the
LDA$\rightarrow$HDA transformation; (iii) we investigate structural evolution
during  the annealing/isobaric heating of the amorphous ices obtained upon
compression of LDA; and (iv) we compare the location in the order map of 
 the various glasses obtained in (i), (ii) and (iii). By doing this, 
we are able to characterize {\it quantitatively} structural differences between
 samples of amorphous ice obtained by different procedures.
 In recent related work, Guillot and Guissani \cite{french} performed computer
simulations of glassy water obtained along  different paths.
 They used the radial distribution function and structure factor
to study the structure of the different glasses.
In this work, we instead use the order map approach. As we show below, this
approach allows a precise and sensitive characterization of structure.

This work is organized as follows. 
In the next Section, we present the simulations details.
In Section~\ref{cooling}, we study the local structure when the liquid is
cooled isochorically across the glass transition at different cooling rates
and for  different densities. We study both the structure in the
instantaneous configurations and in the corresponding potential energy minima
attained by steepest descent mapping (inherent structures, IS).
 In Section~\ref{compressionLDA}, we study the
evolution of structural order during the LDA$\rightarrow$HDA
transformation, and upon annealing (i.e., isobaric heating)
 different glasses obtained during the compression procedure. 
The comparison of the local structure of the glasses obtained with 
 the different procedures mentioned above is discussed in
 Section~\ref{comparisonGlasses}. We present the conclusions in
 Section~\ref{conclusions}.

\section{Simulations}
\label{simulations}

We perform classic molecular dynamics (MD) simulations of a ($N=216$)-molecule system 
using the extended simple 
point charge (SPC/E) model of water\cite{berensen}. This model has been
extensively used to study the thermodynamics
\cite{francislong,pooleSPC/E} and dynamics\cite{francescoSPCE2,ourSHD}
of liquid water and is  consistent with experimental facts (e.g., at low
temperatures it exhibits density anomalies \cite{francescoSPCE1} and a
diffusivity that increases upon compression \cite{francislong}).
  The SPC/E model has also been used to
study glassy water, and is able to produce glassy states corresponding to
LDA, HDA and VHDA\cite{ourVHDA,ourLDAHDA}. 

To cool the liquid to the glass state, 
we perform constant volume simulations 
at $\rho=0.9,~1.0,~1.1,~1.2,~1.3$ and $1.4$~g/cm$^3$. At each
density we average the results over 16 independent simulations (with the
exception of $\rho=1.00$~g/cm$^3$ where we use 32 independent simulations).
We use periodic
 boundary conditions and the reaction field
 method (with a cutoff of $0.79$~nm) to treat the long range forces.
Upon cooling at each density,
 we change the $T$ by rescaling the
velocities of the molecules using the Berendsen thermostat
\cite{berendsenThermo}. 
  At every time step $\delta t=1$~fs, we decrease
the thermostat temperature by $\delta T= q_c \delta t$, where $q_c<0$ is the
cooling rate.
Cooling simulations starting from an
equilibrium liquid at $T=300$~K down to $T=0$~K are performed with cooling
rates  $q_c=-30,~-10^2,~-10^3,~-10^4$ and $-10^5$~K/ns.
Similar values for $q_c$ have been
used in previous simulations at $\rho=1.00$~g/cm$^3$ 
to study the effect of cooling and heating rates on the glass transition
temperature $T_g=T_g(P)$ \cite{ourSHADOW,ourPRE-RC}.  The
glass obtained with a cooling rate $q_c=-30$~K/ns behaves,
 upon heating at $q_h=+30$~K/ns, as an experimentally slow-cooled glass, and
shows no signs of hyperquenching effects \cite{ourPRE-RC}.
 On the other hand, the smallest cooling/heating rates accessible nowadays in computer
 simulations  are $|q| \approx 10$~K/ns \cite{french}.

To study the evolution of structural order during the LDA$\rightarrow$HDA transformation, 
we perform compression simulations at $T=77$ and $170$~K and average results over 16
independent trajectories at each temperature.
 We perform MD simulations at constant
 $\rho$ for intervals of $1$~ps and at the end of each interval we increase $\rho$ by 
$\delta \rho =  5\times 10^{-5}$~g/cm$^3$. These $\rho$ changes are
performed by  rescaling isotropically
 the coordinates of the molecular center of mass. Therefore, our 
compression rate is $\delta \rho / \delta t=  5\times
10^{-5}$~g/cm$^3$/ps.  This value was already used to study the potential
energy landscape region sampled during
 the LDA-HDA transformation\cite{ourLDAHDA}. 
We also perform 16 independent compression simulations at $T=0$~K;
 at each simulation step  we change $\rho$ by $\delta \rho = 5\times 10^{-5}$~g/cm$^3$ and
 then minimize the energy. At each step, $\rho$ is modified by  rescaling
 isotropically the center of mass of each molecule. 
 
We stress that our results are relative to the cooling/compression rates
studied here. However, the cooling/compression rates we use are such that 
the compression run ($0.9 \leq \rho \leq 1.4$~g/cm$^3$)
 and the slowest cooling run ($300 \geq T \geq 0$~K)
both take $10$~ns, i.e. both numerical `experiments' are based on the same
time scale (as is expected to be the case in an experiment). 

The annealing procedure of the glasses obtained upon compression consists of 
isobaric MD simulations where the temperature is increased from $T=77$~K
(i.e., the compression temperature used in the  LDA$\rightarrow$HDA transformation)
 up to $T=170$~K. We use the Berendsen thermostat
\cite{berendsenThermo} to fix the $T$, and at every time step $\delta t=1$~fs, we increase
the thermostat temperature by $\delta T= q_h \delta t$, where $q_h=+30$~K/ns is the
heating rate. A coupling to an external bath at a given  $P$ (analogous to
the Berendsen thermostat) is used to keep the pressure constant \cite{berendsenThermo}.
We perform annealing runs at
$P=-0.55,~-0.17,~0.01,~0.41,~0.68,~0.84,~1.10,~1.38$ and $1.90$~GPa
 and average results over 16 trajectories.
 
We also investigate structural order in mechanically stable configurations
 sampled during isochoric cooling. These configurations are
 local potential energy minima called inherent structures (IS).
To obtain the IS we perform energy minimizations by using the conjugate
gradient algorithm. We consider the minimization
complete when the change in energy between two successive minimizations 
is $ \leq 10^{-15}$~kJ/mol.

We compute for the glasses the same translational and orientational order
 parameters $\tau$ and $Q$ introduced in
\cite{jeff} to study structural order in liquid water.
The translational order parameter $\tau$ is given by
\begin{eqnarray} 
\tau \equiv \frac{1}{\xi_c} \int_0^{\xi_c} |g_{OO}(\xi)-1| d\xi
\label{tauDef}
\end{eqnarray} 
where $\xi \equiv r \rho_n^{1/3}$ is the distance between the oxygen atoms of pairs
of molecules, $r$, divided by the mean nearest-neighbor separation
at the number density $\rho_n \equiv N/V$; $g_{OO}(r)$ is the oxygen-oxygen radial
distribution function (RDF) and $\xi_c \equiv 2.843$ is a cut-off distance.
In an ideal gas, the RDF is equal to $1$ and 
$\tau =0$. In a crystal,  there is long-range order and $g_{OO}(r) \neq 1$
over long distances, and $\tau$ is large.

The orientational order parameter $Q$ is given by
\begin{eqnarray} 
Q \equiv 1 - \frac{3}{8} \sum_{j=1}^3 \sum_{k=j+1}^4 \left( \cos \psi_{jk} +
  \frac{1}{3}    \right)^2 
\label{Qdef}
\end{eqnarray} 
where $\psi_{jk}$ is the angle formed by the lines joining the oxygen atom of
a given molecule and those of nearest neighbors $j$ and $k$ ($\leq 4$).
For the purpose of this calculation, we limit our attention to the four
oxygen atoms that are closest to a given oxygen atom.
This definition is a slightly modified version of the metric  introduced in
Ref.~\cite{chau}. Equation~(\ref{Qdef}) implies that $-3\leq Q \leq 1$.
In an ideal gas, $<Q>=0$ (where $<...>$ denotes an ensemble average) \cite{jeff}. 
In a perfect tetrahedral network $\cos(\psi_{jk}) = -1/3$, and $Q=1$. Thus,
$Q$ measures the degree of tetrahedrality in the distribution of the four
nearest oxygen atoms around a central oxygen atom.

\section{Structural Order upon  Isochoric  Cooling}
\label{cooling}

\subsection{Instantaneous Configurations}

\begin{figure}[p]
\narrowtext 
\centerline{
\hbox {
  \vspace*{0.5cm}  
  \epsfxsize=9cm
  \epsfbox{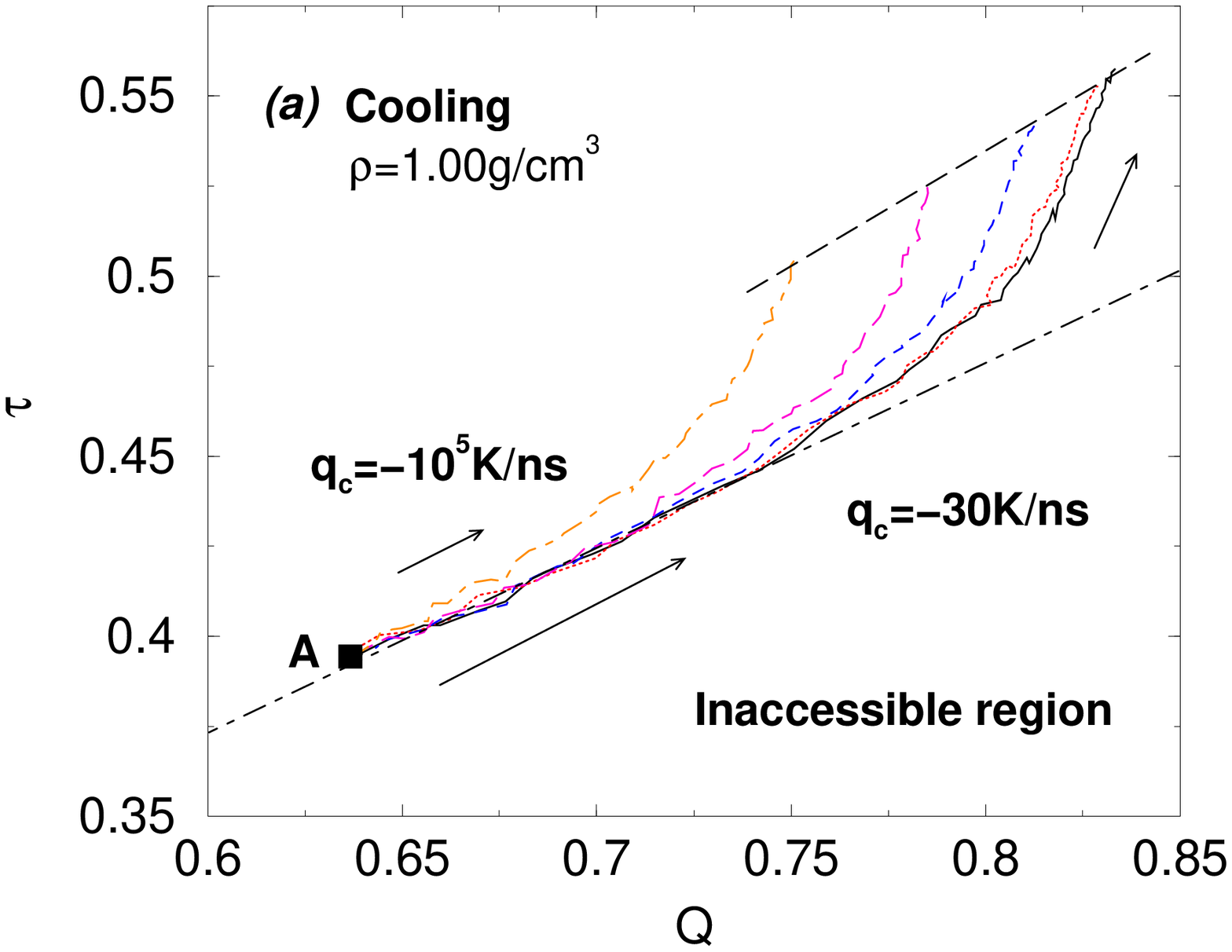}
}}
\centerline{
\hbox {
  \vspace*{0.5cm}  
  \epsfxsize=9cm
  \epsfbox{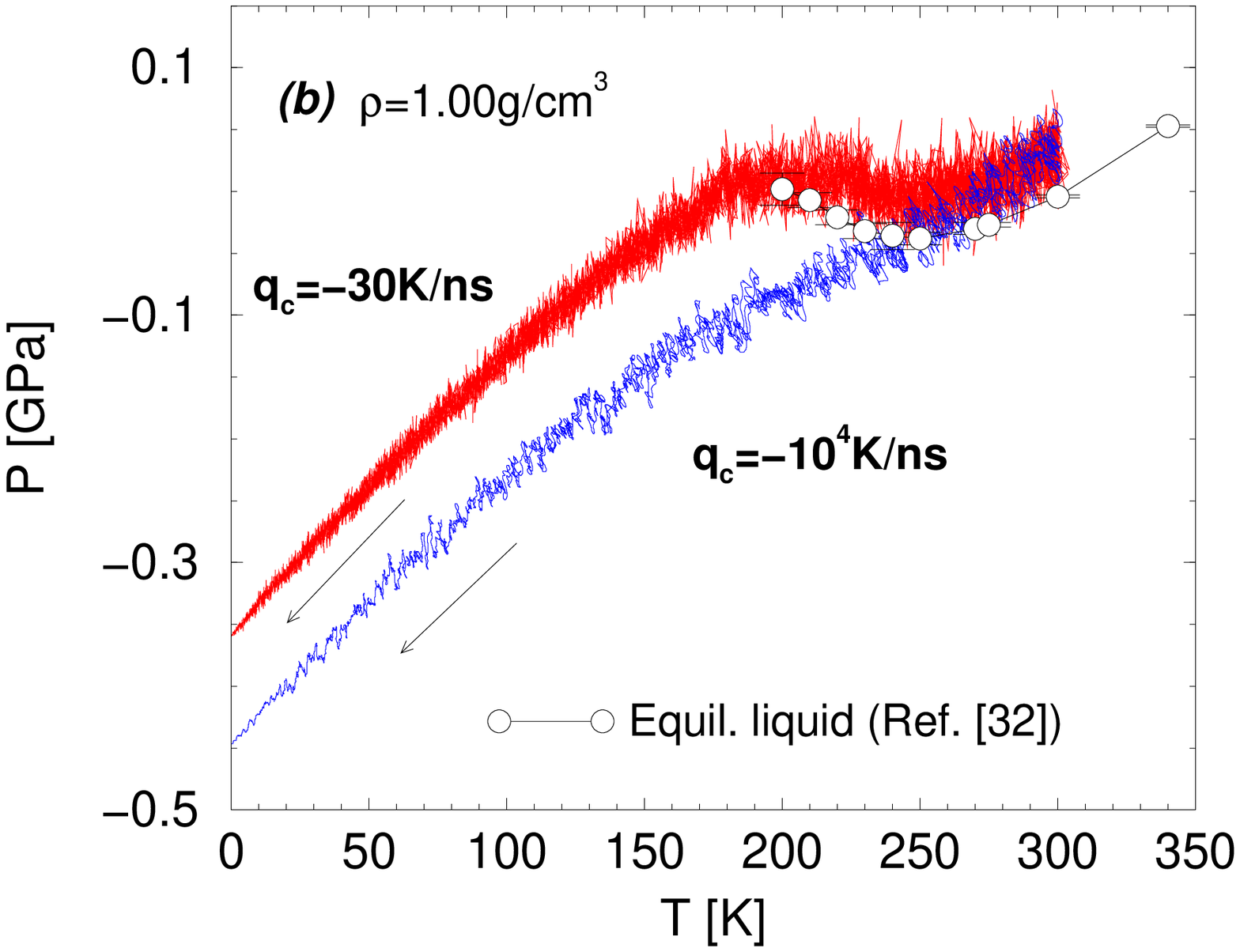}
}}
\centerline{
\hbox {
  \vspace*{0.5cm}  
  \epsfxsize=9cm
  \epsfbox{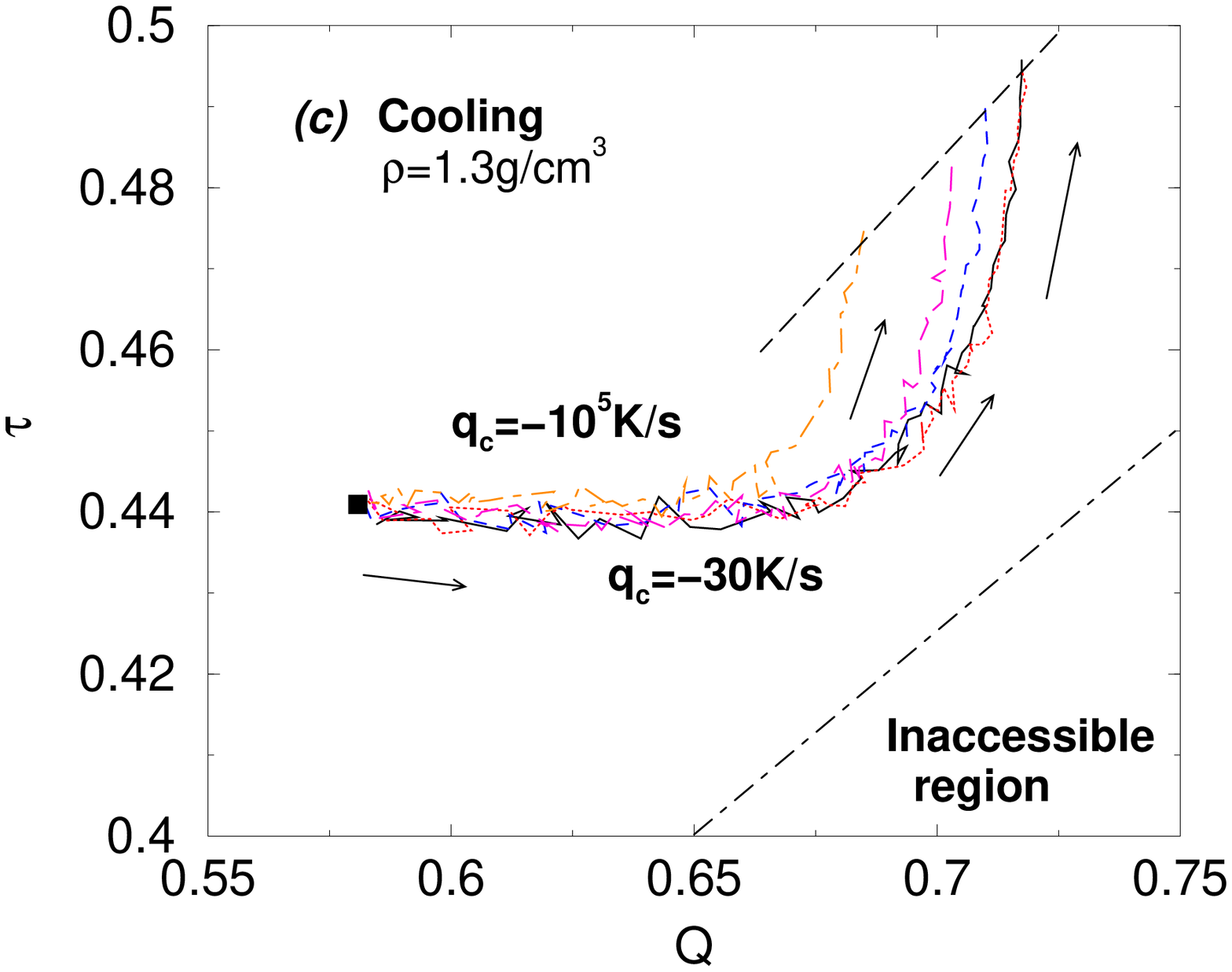}
}}
\vspace*{0.5cm}
\caption{(a) Behavior of the orientational and translational 
order parameters ($Q$ and $\tau$, respectively) 
 when cooling an initially equilibrium liquid (point `A': $T=300$~K,
 $\rho=1.0$~g/cm$^3$) at constant density
 $~\rho=1.00$~g/cm$^3$. Upon cooling at the slower rate, $q_c=-30$~K/ns,
 the system initially 
moves along the path corresponding to equilibrium $Q-\tau$ values 
(in (a), this path
coincides with the dot-dashed line delimiting the inaccessible region for the
liquid state). At lower $T$, the system deviates from the equilibrium path.
The long-dashed line is the result of interpolating the
$Q$-$\tau$ values of the glasses obtained at $T= 0$~K at different $q_c$.
 The various cooling rates correspond to $q_c$ values of 
$-10^5,~-10^3,~-10^3,~-10^2$ and $-30$~K/ns (left
 to right).
(b) Pressure as a function of $T$ upon cooling the liquid at
$\rho=1.00$~g/cm$^3$ for two different cooling rates.
(c) Same as (a) for $~\rho=1.30$~g/cm$^3$.}
\label{qt-plane}
\end{figure}

Figure~\ref{qt-plane}(a) shows the evolution of the order parameters $Q$ and
$\tau$ during cooling of an initially equilibrated liquid at constant density
$~\rho=1.00$~g/cm$^3$ from $T=300$~K
down to $T=0$~K. We show the results for different cooling rates $q_c$ and
include the line obtained in \cite{jeff}
separating the accessible and the inaccessible regions for {\it equilibrium
  liquids}. No equilibrium liquid states of the SPC/E model exist in the
range $220$~K$\leq T \leq 400$~K and $0.85$~g/cm$^3$$\leq \rho \leq
1.3$~g/cm$^3$ with order parameter combinations $(Q, \tau )$ to the right of
this limiting line.

 The starting location of the system in the ordering map is
indicated by the square symbol `A'
 and is in agreement with the values reported in \cite{jeff} for a
 liquid at $T=300$~K and $~\rho=1.0$~g/cm$^3$. 
Upon cooling at the slower rate $q_c=-30$~K/ns, the system
moves in the order map along
the line delimiting the accessible region for the liquid state and finally,
at approximately $T= 240$~K, the system departs from this line and 
moves into the
accessible region. The boundary between the accessible and inaccessible regions
is the locus of state points where structural, dynamic, and
thermodynamic anomalies occur \cite{jeff}. 
Equilibrium liquid states that display structural, dynamic, or thermodynamic 
anomalies (decrease of both $\tau$ and $Q$ upon isothermal compression; 
increase in diffusivity upon isothermal compression; 
decrease in density upon isobaric cooling, respectively) are therefore
characterized by only one order parameter ({\it line} in the $Q$-$\tau$
plane). In contrast, liquid states that do not exhibit anomalies are
characterized by two order parameters (accessible {\it area} in the $Q$-$\tau$
plane) \cite{jeff}.
Therefore, at $\rho = 1.0$~g/cm$^3$ and $300$~K~$\geq T \geq 240$~K, 
the equilibrium liquid experiences a cascade of anomalies (first structural,
then dynamic, finally thermodynamic) upon cooling \cite{jeff}.
For $T < 240$~K, the quenched liquid falls out of equilibrium and in so doing 
moves into the accessible region of the order map. We use the term
equilibrium to denote states where the structural relaxation time is short
compared to the simulated time at the given $T$ and $\rho$. Equilibrium
states for the liquid at $\rho=1.0$~g/cm$^3$ and $220$~K$\leq T \leq 300$~K lie on the
limiting line separating the accessible and inaccessible regions \cite{jeff}.

In order to understand the trajectory followed by the quenched system at 
$\rho = 1.00$~g/cm$^3$ in the order map we note that  the liquid is
able to reach equilibrium during cooling down to $T \approx 240$~K. 
The structural relaxation time at $T>240$~K is $t_r<30$~ps
\cite{francislong} while during 
the cooling process, the typical time scale for the
system to relax before $T$ changes by $1$~K is $dt=1$~K$/|q_c|\approx
33$~ps~$\approx t_r$. For $T<240$~K, therefore, 
 the system is not able to reach equilibrium
 upon cooling (i.e., $dt<t_r$) and it attains, for a given $Q$, larger
 values of $\tau$
 than those possessed by the liquid in equilibrium at the given $T$ and
 $\rho$.
 Equivalently, the quenched system attains smaller values of $Q$, for a given
 $\tau$, than those of the equilibrium liquid at the given $T$ and $\rho$.

It is interesting to study the behavior of the thermodynamic properties 
when the system falls out of equilibrium upon cooling. 
 Figure~\ref{qt-plane}(b) shows the behavior of pressure
 $P$ as a function of $T$
 upon cooling at $q_c=-30$~K/ns and $q_c=-10^4$~K/ns.
As a guide to the eye, we also include data from equilibrium simulations taken 
from Ref.~\cite{pooleSPC/E}. Upon cooling at $q_c=-30$~K/ns the
pressure clearly follows the equilibrium $P(T)$ trend, showing a
minimum at $T\approx 250$~K. A minimum in $P(T)$ at constant density
 corresponds to a maximum in $\rho(T)$ at constant density
 (see e.g. \cite{francislong}).
 Therefore, upon cooling at $q_c=-30$~K/ns, the liquid is able to 
experience a density anomaly 
[$\left( \partial P / \partial T \right)_\rho < 0$] in the approximate
range $250$~K$ \geq T \geq 190$~K, whereas for the equilibrium liquid, density
 anomalies occur for all $T\leq 250$~K.
 When the liquid is cooled faster, at $q_c=-10^4$~K/ns, the
pressure deviates from the equilibrium $P(T)$ curve at a comparatively 
higher $T$. No signature
of a density anomaly is observed when cooling at this high rate.
This suggests that configurations responsible for producing density anomalies
possess comparatively larger equilibration times. This observation warrants a
systematic investigation. 
We note that pressure fluctuations make it difficult to estimate the $T$ at
which the instantaneous $P$ deviated from the equilibrium $P(T)$ curve.

The phenomenology discussed above when relating the behavior of the order
parameters upon cooling at $q_c=-30$~K/ns is common
to all the cooling rates studied [see Fig.~\ref{qt-plane}(a)].
 However, we observe that the larger
$|q_c|$, the higher the $T$ at which the system falls out of equilibrium (i.e., 
the earlier the system deviates from the line delimiting the
inaccessible region). Furthermore, 
the larger is $|q_c|$, the closer are the final values of
$Q$ and $\tau$ to those of  the starting liquid (square symbol `A' in the
figure), meaning 
that the structure of the glass is more similar to that of
the starting liquid. Equivalently, the smaller is $|q_c|$, the more structured
is the final glass, i.e., the larger the final values of $Q$ and $\tau$.

The long-dashed line in Fig.~\ref{qt-plane}(a) is
 the locus of $Q$-$\tau$ values
 for $\rho = 1.0$~g/cm$^3$ glasses obtained at $T= 0$~K with
the five $q_c$ studied.
 Interestingly, such isochorically-quenched 
glasses define a line in the order map.
We note that these results are similar to simulations performed
using a Lenard-Jones potential \cite{qt-LJ}. In fact, Fig.~7 in that work
shows that upon isochorically cooling a liquid with different cooling rates,
 the orientational and
translational parameters increase monotonically with decreasing cooling rate,
 and the glasses obtained at the lowest $T$ fall on a straight line with positive
slope in the order map, just as we find for SPC/E water. 

The features observed at $\rho=1.00$~g/cm$^3$ are also found at all the
other densities studied, i.e., $\rho=0.9,~1.1,~1.2,~1.3$, and $1.4$~g/cm$^3$.
The results for $\rho=1.3$~g/cm$^3$ are shown in Fig.~\ref{qt-plane}(c).
In agreement with \cite{jeff}, comparison between Figs.~\ref{qt-plane}(a) and
(c) indicates that the location in the order map of the starting
 liquid at $T=300$~K (square symbol) depends on the density.
Furthermore, at the high density $\rho=1.3$~g/cm$^3$ the equilibrium liquid
does not exhibit structural, dynamic, or thermodynamic anomalies.
Accordingly, the quench trajectories in the order map do not attain the line
separating the accessible and inaccessible regions.
We note that at all the densities and cooling rates studied, the glasses
 always remain out of the inaccessible region obtained previously for
 the equilibrium liquids \cite{jeff}. In this
respect, the inaccessible region in the order map seems to be the
 result of topological  constrains inherent 
in the molecular interactions and  does not depend on 
whether the system is in the liquid or glassy state.

\begin{figure}[p]
\narrowtext 
\centerline{
\hbox {
  \vspace*{0.5cm}  
  \epsfxsize=12cm
  \epsfbox{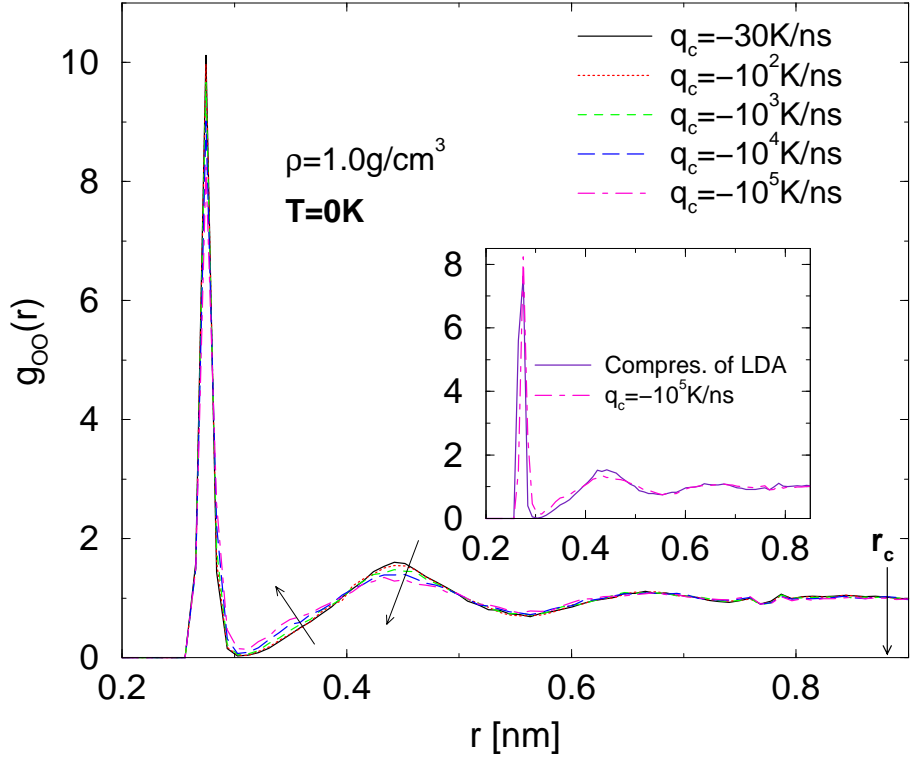}
}}
\vspace*{0.5cm}
\caption{ Radial distribution function
 $g_{OO}(r)$ corresponding to the glasses at $T=0$~K prepared by isochoric
  cooling at $\rho=1.00$~g/cm$^3$ and different cooling rates.
As the cooling rate increases (see arrows), mild changes in 
 the first minimum and the second
maximum of $g_{OO}(r)$ are observed indicating that 
 molecules in the second shell move
closer to the central molecule.
These glasses show a very similar $g_{OO}(r)$ but have quite different order
parameters [see Fig.~\ref{qt-plane}(a)]. Same conclusion holds when
comparing $g_{OO}(r)$ for the glass
 obtained upon $T=0$~K-isothermal compression at $\rho=1.0$~g/cm$^3$ and
for the isochorically quenched glass at $q_c=-10^5$~K/ns at the same density
and temperature (see inset). The cutoff value $\xi_c \equiv r_c \rho_n^{1/3}$
used in the definition of $\tau$ (see Eq.~\ref{tauDef}) corresponds to a cutoff
distance $r_c= 0.883$~nm at $\rho=1.0$~g/cm$^3$. The value of $r_c$ is
indicated by the arrow in the lower right corner.
}
\label{rdf-cooling}
\end{figure}

The radial distribution function is one of the standard tools used in
experiments, theory,
 and simulations to characterize the structure of condensed matter.
 We show in Fig.~\ref{rdf-cooling} the oxygen-oxygen radial distribution
function $g_{OO}(r)$ for the glasses obtained at $T=0$~K upon cooling at
$\rho=1.00$~g/cm$^3$ with the same five cooling rates shown in
 Fig.~\ref{qt-plane}(a). These 
glasses are located in the order map along the long-dashed line.
While differences in the structures of the glasses are suggested by both 
Fig.~\ref{qt-plane}(a) and Fig.~\ref{rdf-cooling}, the differences observed
in $g_{OO}(r)$ are quite subtle and could be difficult to distinguish from
typical experimental error 
bars. It therefore appears that the location of the system in the 
$(Q, \tau)$ {\it order map} is a sensitive indicator of glass structure.
Same conclusion holds when comparing $g_{OO}(r)$ 
for the glasses obtained upon compression of LDA at $T=0$~K
 (where no annealing is possible) and the isochorically fast-quenched
 glasses at the same $T$. For example, 
  inset of Fig.~\ref{rdf-cooling} compares the $g_{OO}(r)$ for the glass
 obtained upon isothermal compression at $T=0$~K to $\rho=1.0$~g/cm$^3$ and
for the isochorically quenched glass at $q_c=-10^5$~K/ns at the same density
and temperature. The
 glasses obtained by isothermal compression and those prepared by other routes 
 will be discussed and compared in detail in the rest of the paper. The point
 here, though, is to emphasize that the differences in  $g_{OO}(r)$ shown in
 the inset of Fig.~\ref{rdf-cooling}  are quite
subtle, whereas the two glasses have very different order metrics:
$Q=0.83$, $\tau=0.57$ for $T=0$~K-compression of LDA; $Q=75$, $\tau=50$ for 
the   isochorically-quenched glasses at $q_c=-10^5$~K/ns.

\subsection{Inherent Structures}
\label{coolingIS}

We now investigate the location in the order map of 
 the IS sampled by the system upon cooling from $T=300$~K down
to $T=0$~K. We denote the tetrahedral and translational order parameter of
 the IS by $Q_{IS}$ and $\tau_{IS}$, respectively. 

\begin{figure}[p]
\narrowtext 

\centerline{
\hbox {
  \vspace*{0.5cm}  
  \epsfxsize=8cm
  \epsfbox{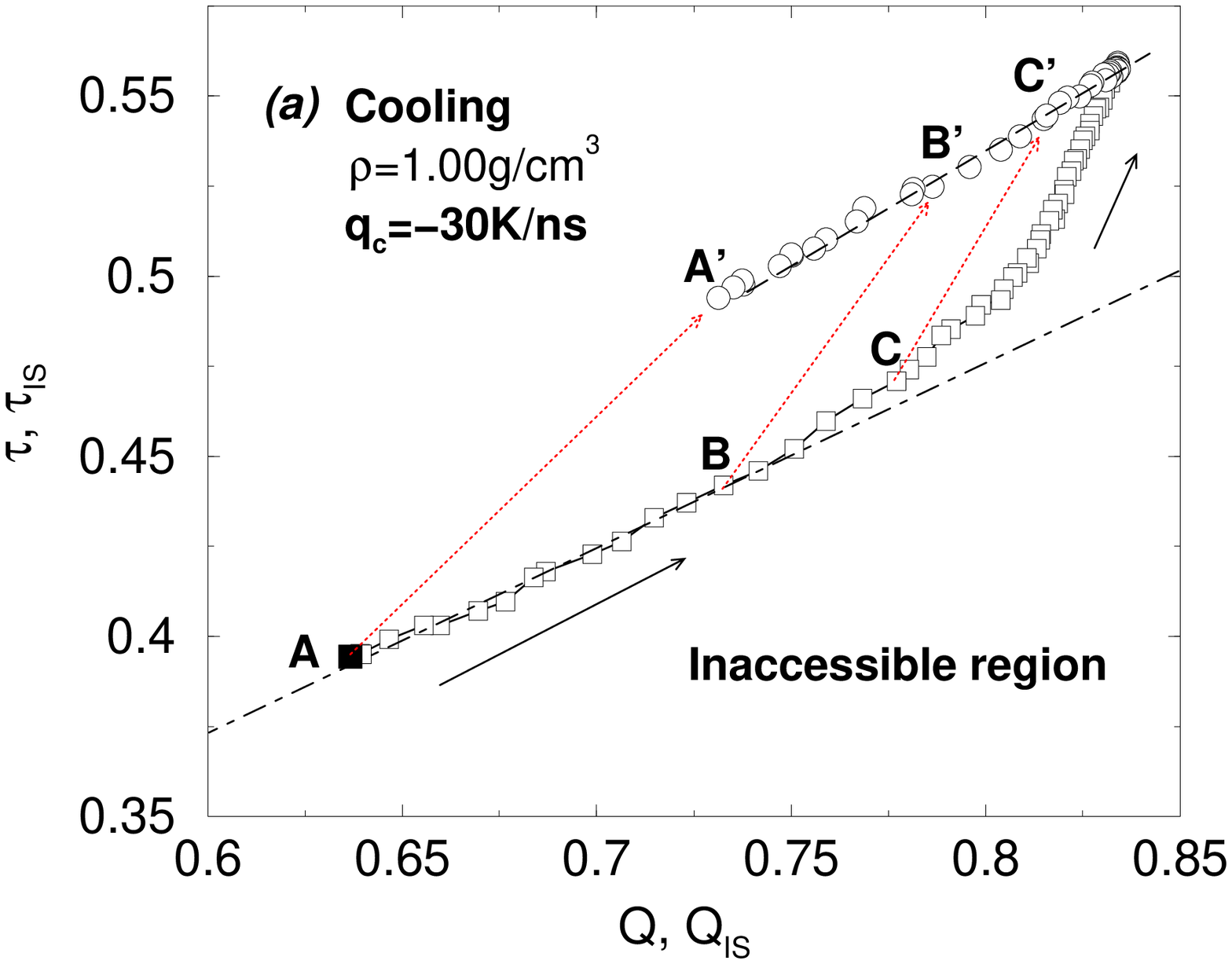}
}}
\centerline{
\hbox {
  \vspace*{0.5cm}  
  \epsfxsize=8cm
  \epsfbox{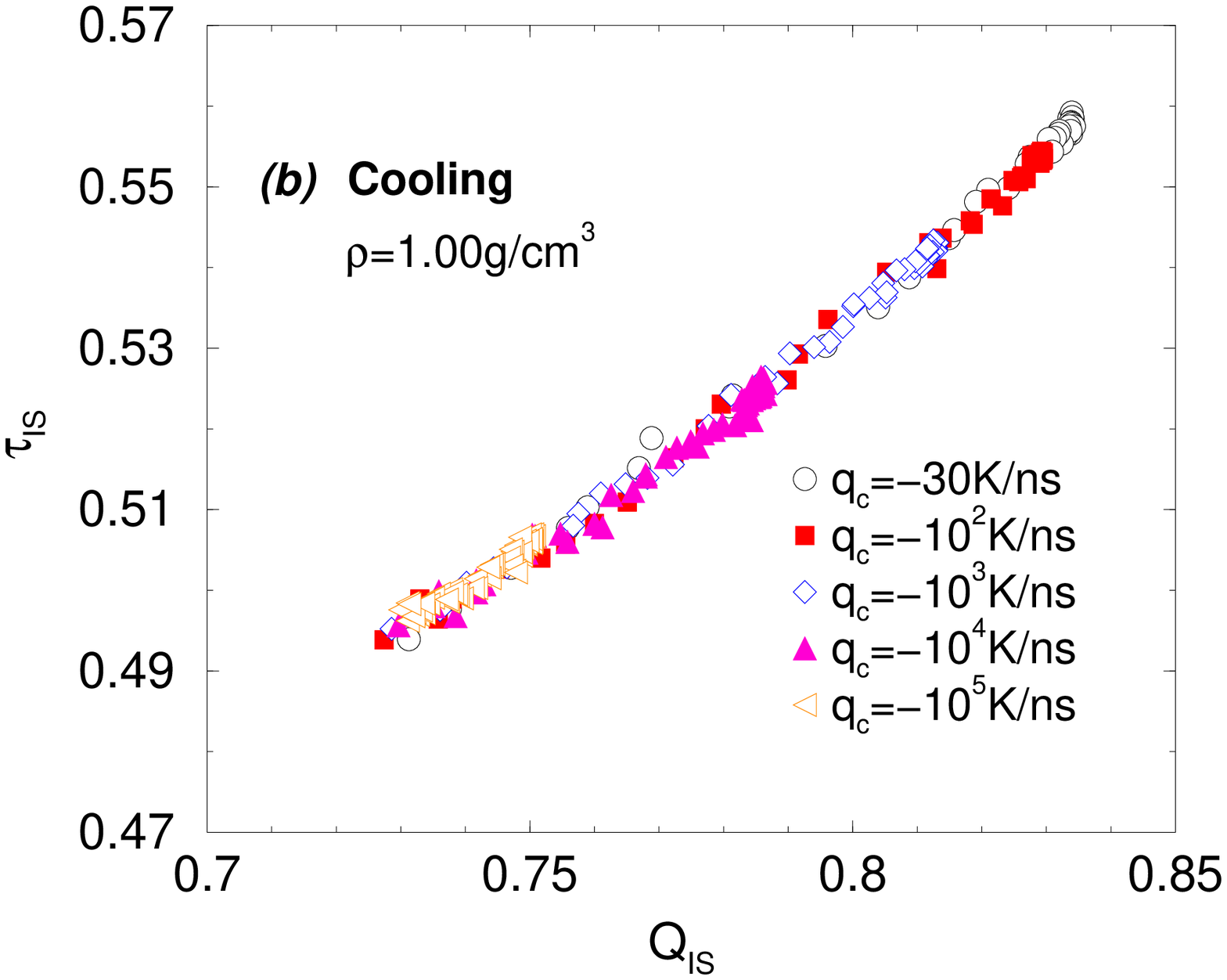}
}}
\centerline{
\hbox {
  \vspace*{0.5cm}  
  \epsfxsize=8cm
  \epsfbox{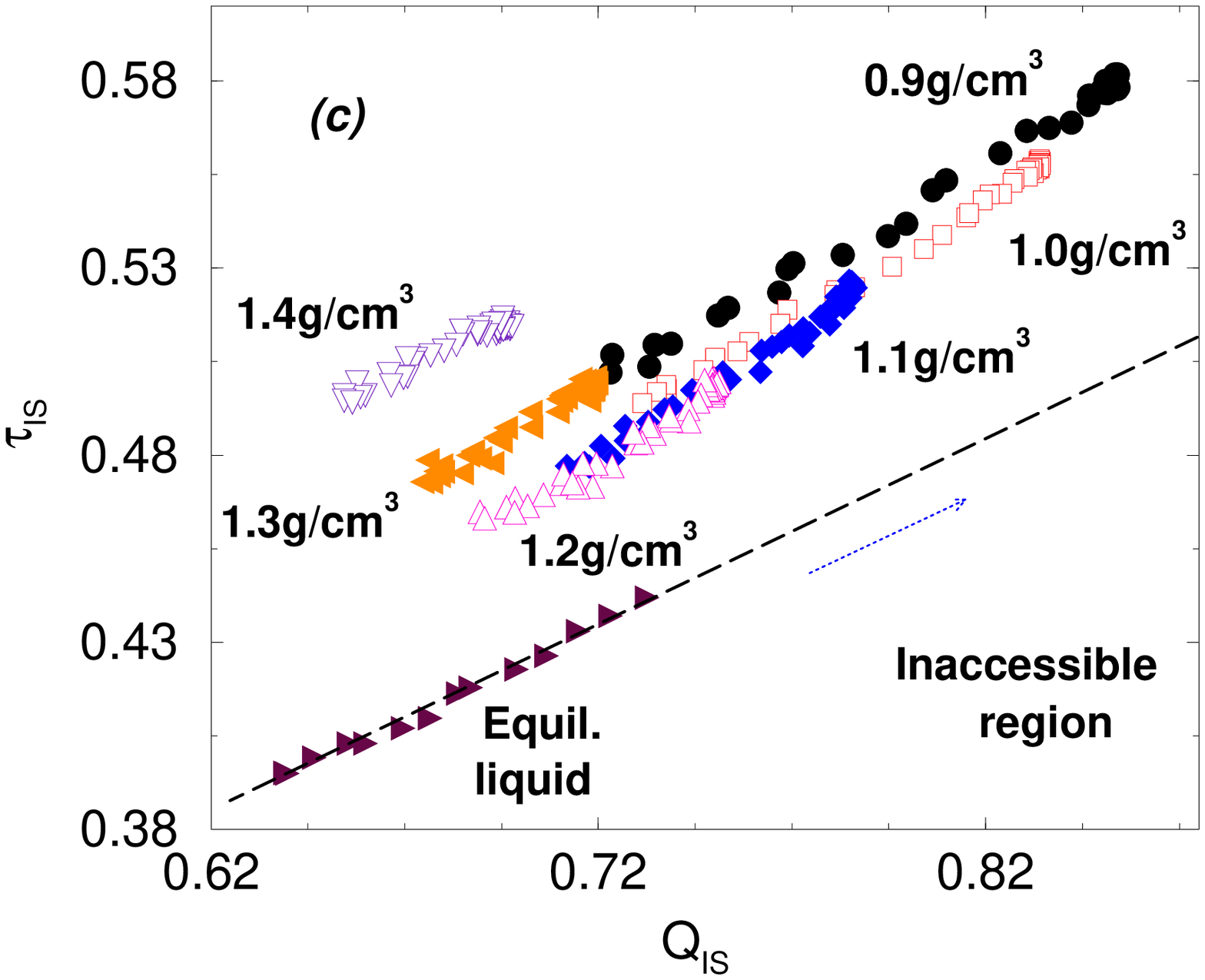}
}}
\centerline{
\hbox {
  \vspace*{0.5cm}  
  \epsfxsize=8cm
  \epsfbox{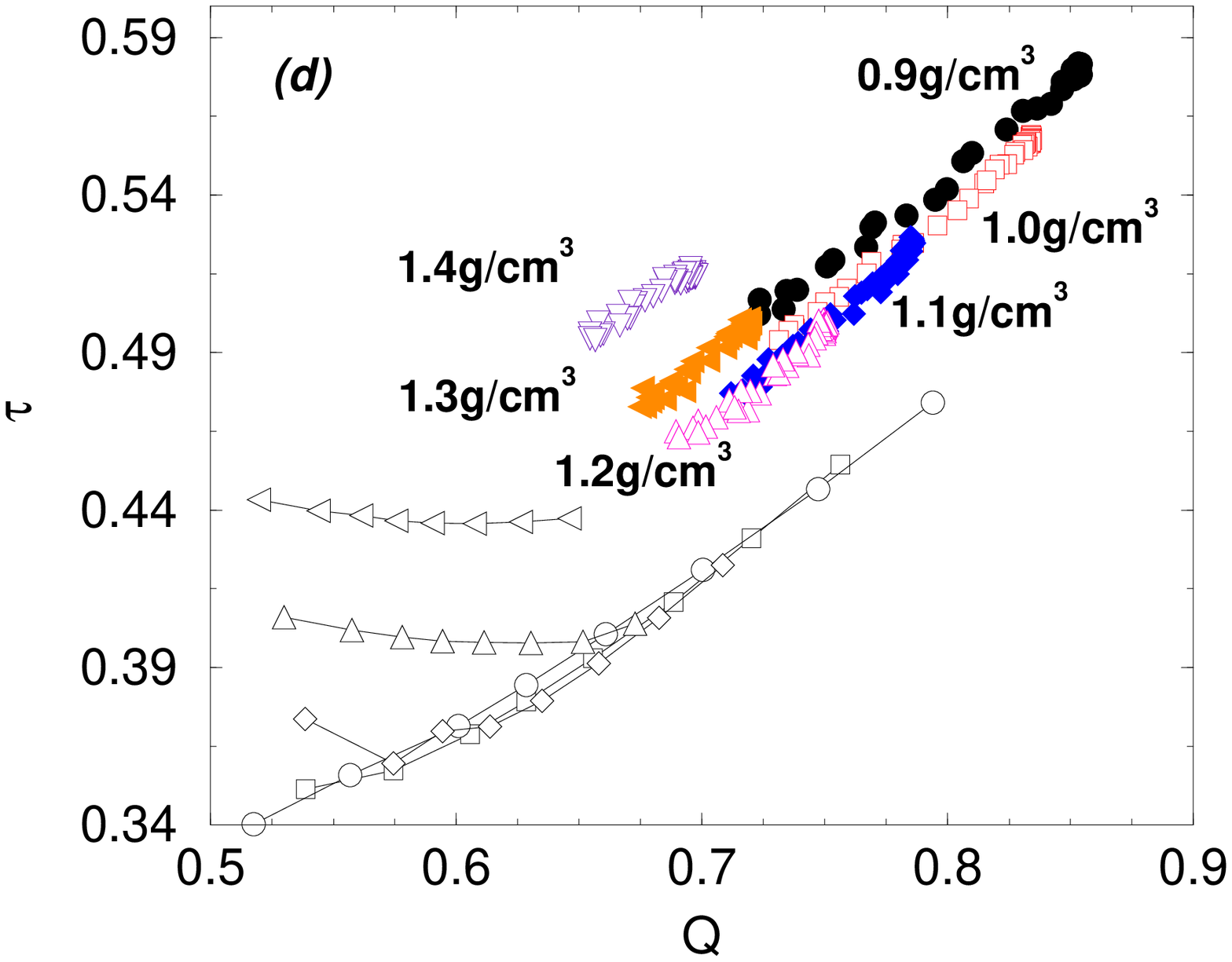}
}}
\vspace*{0.5cm}
\caption{(a) Order parameters $Q$ and $\tau$ for the slowest
  cooling rate shown in Fig.\ref{qt-plane}(a) (square symbols).
 For each configuration obtained upon cooling
 (e.g., points A, B, and C), we obtain
  the corresponding IS (points A', B', and C'). The order parameters in the
  IS ($Q_{IS}$ and $\tau_{IS}$), coincide with the long-dashed line obtained
  in Fig.\ref{qt-plane}(a) for the glasses quenched at $T=0$~K. (b) Location in the order map of the IS sampled
  upon cooling at different cooling rates. All IS fall on the same
  long-dashed line shown in (a). (c) 
Comparison of the location in the order map
  of the IS sampled at different densities at the slowest cooling rate
  studied, $q_c=-30$~K/ns. There is a line characterizing the location of the
  IS in the order map at each $\rho$. 
This line shifts in the order map non-monotonically
  with $\rho$. (d) Comparison between the IS lines shown in (c) and the
 lines obtained by the equilibrium liquid state points at
  different $\rho$ {\protect \cite{jeff}}
(shown at the lower left corner).
At each density, the shape of the symbols (e.g., left-pointing triangles at
 $\rho=1.3$~g/cm$^3$) is the same for the IS and the equilibrium liquid.}
\label{qtIS}
\end{figure}

Figure~\ref{qtIS}(a) shows the evolution of $Q$ and $\tau$ 
upon cooling the liquid at $\rho=1.00$~g/cm$^3$ at $q_c=-30$~K/ns [taken from
Fig.~\ref{qt-plane}(a)]. For each configuration  obtained from
the MD simulations, we find the corresponding IS and then calculate the order
parameters $Q_{IS}$ and $\tau_{IS}$. The evolution of $Q_{IS}$ and
$\tau_{IS}$ upon cooling at $q_c=-30$~K/ns are also indicated in
Fig.~\ref{qtIS}(a).  
Interestingly, we find that the $Q_{IS}$ and
$\tau_{IS}$ values fall on the same line that characterizes the 
glasses obtained  
upon cooling isochorically to $T=0$~K at different $q_c$ [long-dashed line in Fig.~\ref{qt-plane}(a);
included also in Fig.~\ref{qtIS}(a)].
In fact, we find that for all the IS obtained at $\rho=1.00$~g/cm$^3$ 
from systems quenched at {\it any} quenching rate $q_c$,
 the parameters $Q_{IS}$ and $\tau_{IS}$ fall on
the same line that characterizes the isochorically-quenched glasses
at $T \approx 0$~K [see Fig.~\ref{qtIS}(b)].
We also note that the values of the order parameters are larger in the IS
than in the liquid (i.e., the IS are more structured than the liquid configurations).

In Fig.~\ref{qtIS}(c) we show that the behavior
 found at $\rho=1.00$~g/cm$^3$ occurs at all densities studied. 
 For each density, there is a line in the order
map that characterizes the location of the IS sampled upon cooling, which
coincides with the location of the glasses at $T=0$~K.
The location of each of these lines depends non-monotonically on $\rho$. At low $\rho$, the IS-line is located at high $Q$ and $\tau$. As
density increases, this line moves toward lower values of $Q$ and $\tau$, and
finally, at high $\rho$, it shifts to lower values of $Q$ but larger values
of $\tau$. The range of order parameters sampled by the IS
 shrinks appreciably upon compression within the range of $q_c$ values
 studied.

From Ref.~\cite{jeff}, it is found that
equilibrium liquid states of a given density 
fall on a single curve in the $(Q, \tau)$ plane.
 Therefore, it is interesting to investigate 
 how this equilibrium liquid line compares to
 the corresponding IS line at the same density. This is done in
 Fig.~\ref{qtIS}(d) where we show the location in the order map 
of the equilibrium liquid at different $\rho$ (from Ref.~\cite{jeff}) 
and the lines shown in Fig.~\ref{qt-plane}(c) for the corresponding $\rho$.  
Clearly, for a given $\rho$
 the equilibrium liquid line differs from the corresponding IS line.

The finding that the $Q_{IS}$-$\tau_{IS}$ values fall on a density-dependent 
line in the order map has the following implications:
 (i) $Q_{IS}$ and $\tau_{IS}$ are not independent for a fixed density,
 i.e., one cannot change the average 
orientational order in the IS without changing the
translational order; (ii) the function $\tau_{IS}(Q_{IS})$ is independent of
the cooling rate, i.e., on the glass history; and (iii) the function
$\tau_{IS}(Q_{IS})$ is independent of whether the IS corresponds to a system
 in equilibrium or out of equilibrium.
Conclusions (ii) and (iii) are unusual since expressions 
 relating properties of the IS sampled by the system in equilibrium do
  not necessarily apply to IS sampled by the system out of equilibrium.
  For example, the dependence of basin curvature on IS
 depth has been found to be different in IS sampled by  
equilibrium liquids than in IS obtained from fast-quenched 
 or fast-compressed glasses \cite{ourPRE-RC,mossa}, and to depend on the
 cooling/compression rates. 

\begin{figure}[p]
\narrowtext 
\centerline{
\hbox {
  \vspace*{0.5cm}  
  \epsfxsize=9cm
  \epsfbox{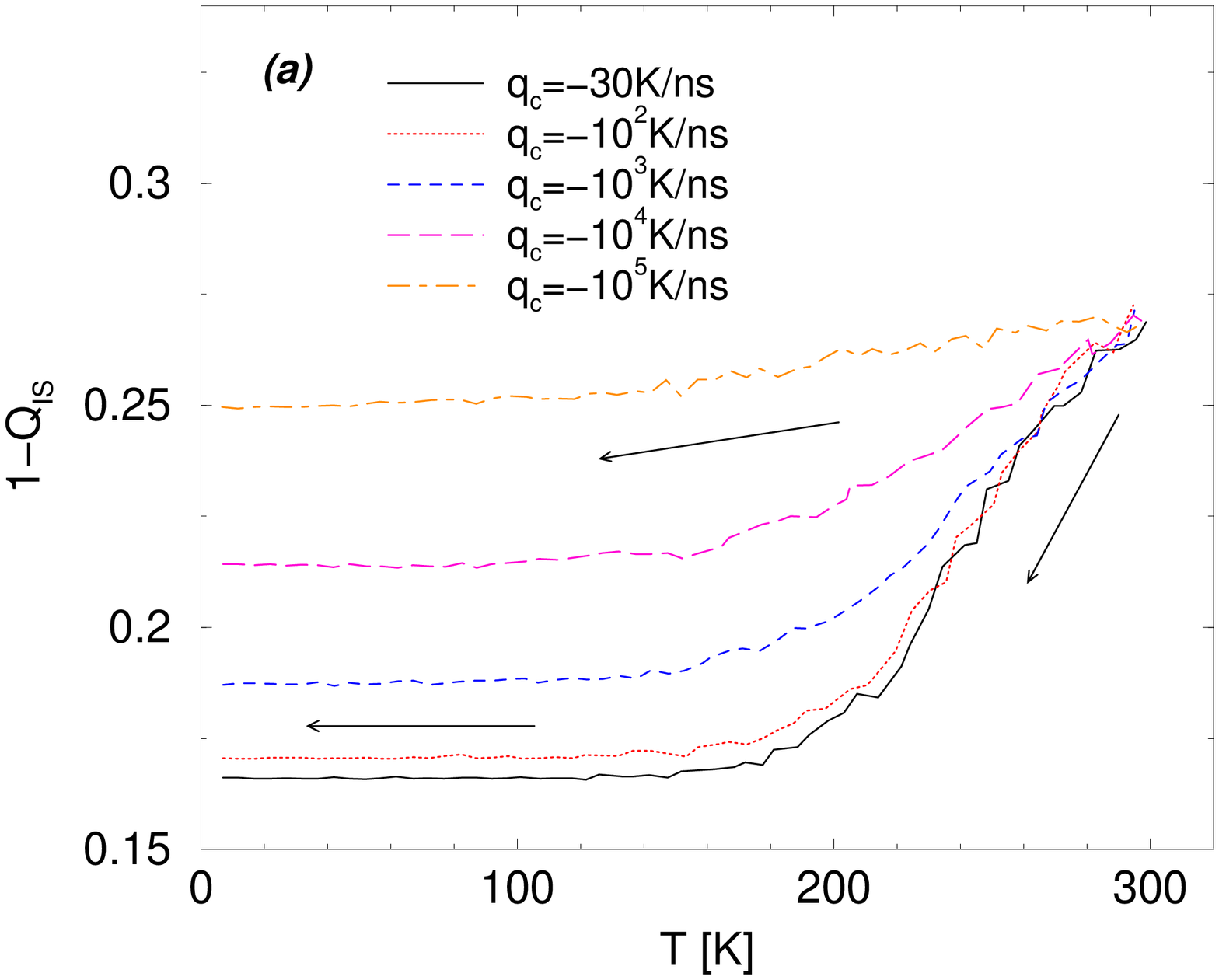}
}}
\centerline{
\hbox {
  \vspace*{0.5cm}  
  \epsfxsize=9cm
  \epsfbox{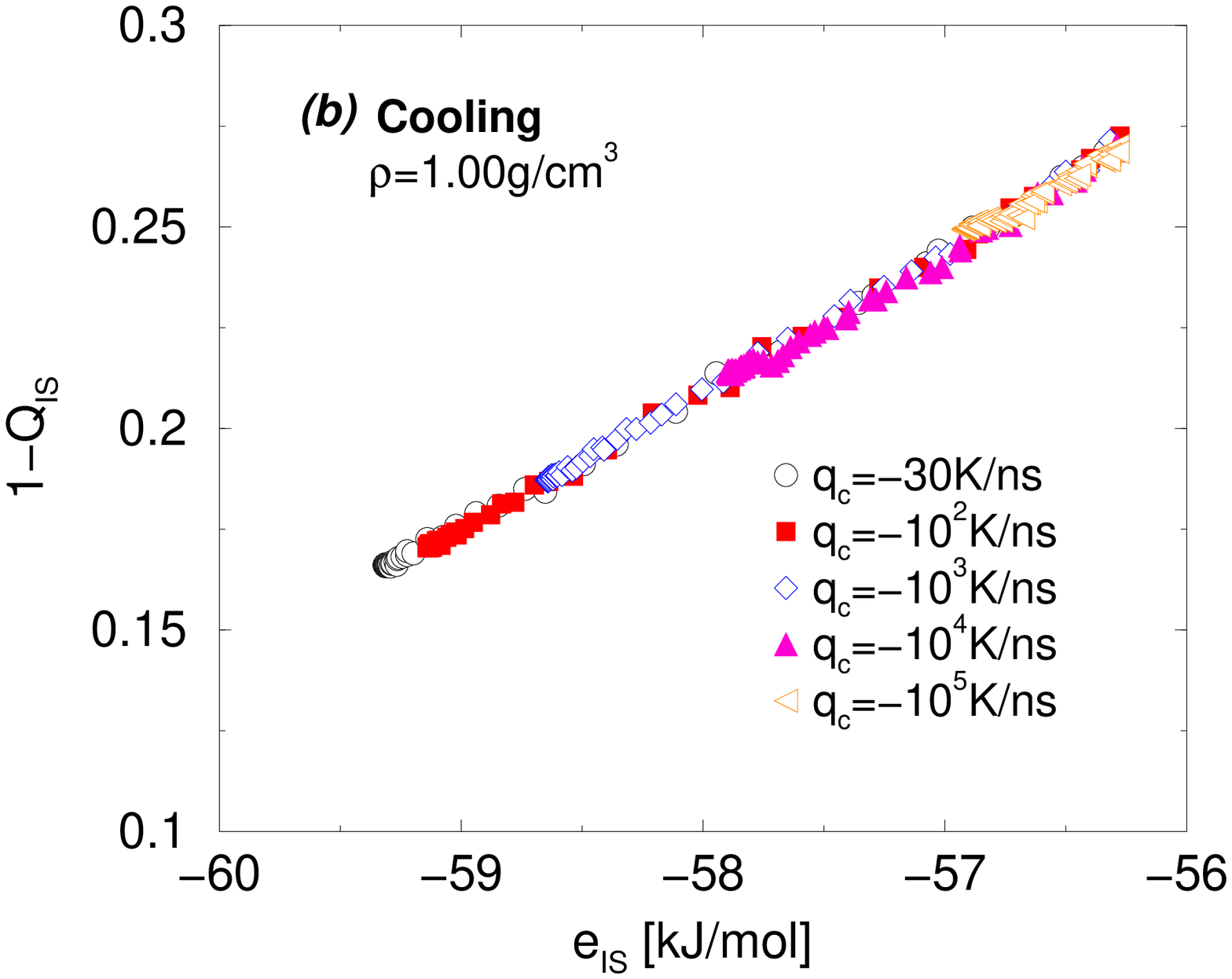}
}}
\vspace*{0.5cm}
\caption{(a) Orientational order parameter $1-Q_{IS}$ in the IS 
as a function of temperature upon cooling at $\rho=1.00$~g/cm$^3$
 at different cooling rates. 
$1-Q_{IS}(T)$ behaves similarly to the energy of the IS observed in 
{\protect \cite{ourPRE-RC}} under same conditions. (b) $Q_{IS}$ 
 as a function of the energy of the IS, $e_{IS}$,  suggesting that 
$1-Q_{IS} \approx e_{IS}$, independently of the cooling rate. Similar results
are found at all densities studied. }
\label{qIS-T}
\end{figure}

In the potential energy landscape (PEL) approach,
 any property of an equilibrium system
\cite{franchescoTheoryPEL} or of a system not too far from equilibrium
 \cite{ourPRE-RC,mossa,kob} is known 
once the energy ($e_{IS}$), local curvature
(${\cal S}_{IS}$), and pressure of the IS ($P_{IS}$) are determined 
\cite{franchescoTheoryPEL}. We investigate the connection between these
properties and the structural order parameters $\tau_{IS}$-$Q_{IS}$.
To do this, Fig.~\ref{qIS-T}(a) shows  $1-Q_{IS}$ as a function of $T$ at
$\rho=1.00$~g/cm$^3$ for all $q_c$ (similar results are obtained for
$\tau_{IS}$).  This figure is reminiscent of the behavior of $e_{IS}$
observed in simulations of a Lenard-Jonnes system \cite{Srinature} and in
water \cite{ourPRE-RC} upon cooling an equilibrium liquid,
 and suggests that $1-Q_{IS} \approx e_{IS}$
 (similarly, we find that $1- \tau_{IS} \approx e_{IS}$) 
meaning that both $Q_{IS}$ and $\tau_{IS}$
are a measure of the basin depth $e_{IS}$ in the PEL. 
 This is confirmed by Fig.~\ref{qIS-T}(b) where we show $1-Q_{IS}$ 
as a function of $e_{IS}$.

\section{Structural Order of Low-density, High-density and  Annealed Amorphous Ice}
\label{compressionLDA}

\subsection{Low-density and High-density Amorphous Ice}

Next, we study the structural order parameters in glassy
water. Specifically, we focus on the values of $Q$ and $\tau$ corresponding
to the glasses obtained during the LDA$\rightarrow$HDA transformation.
 We {\it quantify} the tetrahedrality and translational order 
 of both groups of glasses and identify regions in
 the order map corresponding to LDA and HDA. 
So far, such a characterization has been missing. 

\begin{figure}[p]
\narrowtext 

\centerline{
\hbox {
  \vspace*{0.5cm}  
  \epsfxsize=8cm
  \epsfbox{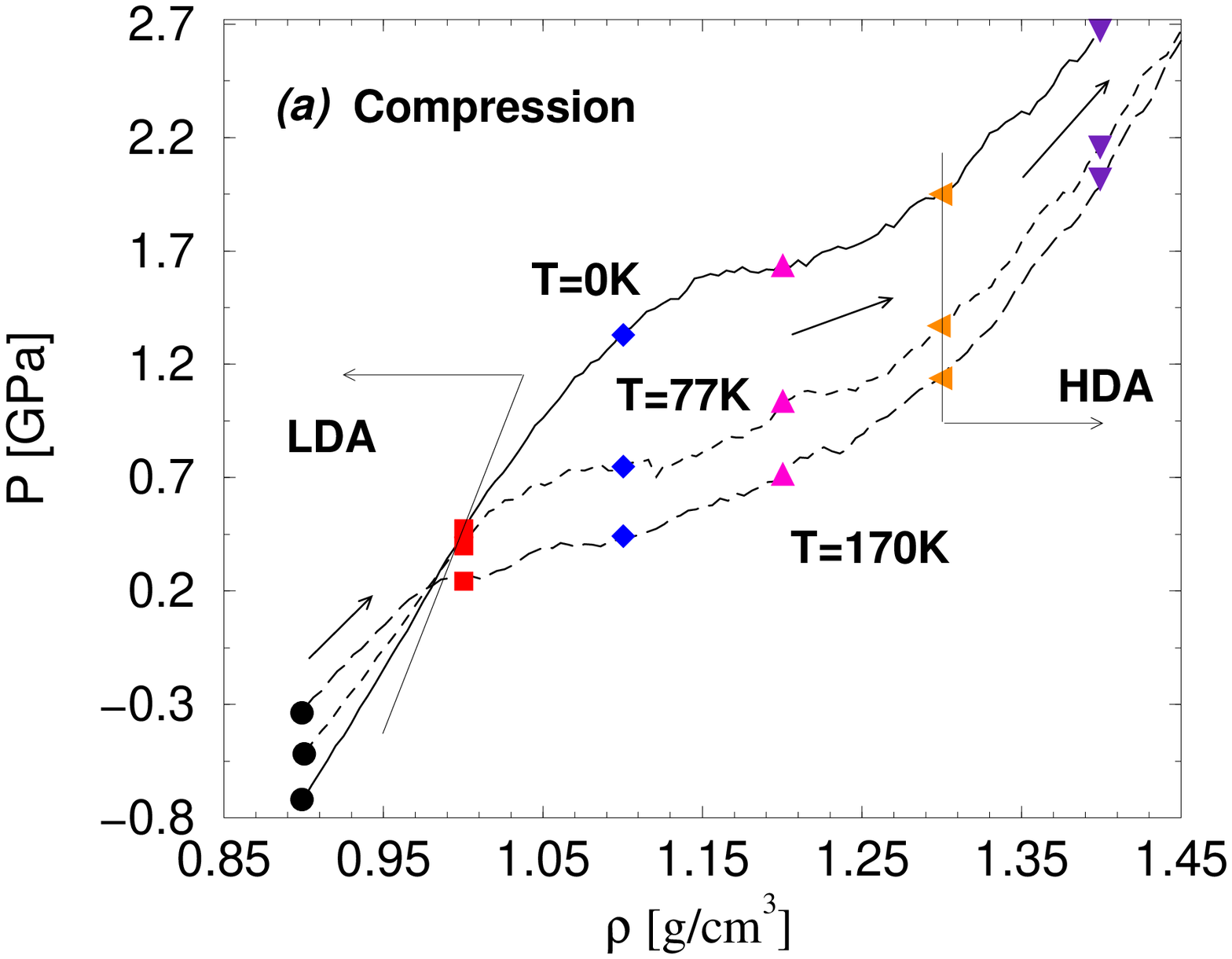}
}}
\centerline{
\hbox {
  \vspace*{0.5cm}  
  \epsfxsize=8cm
  \epsfbox{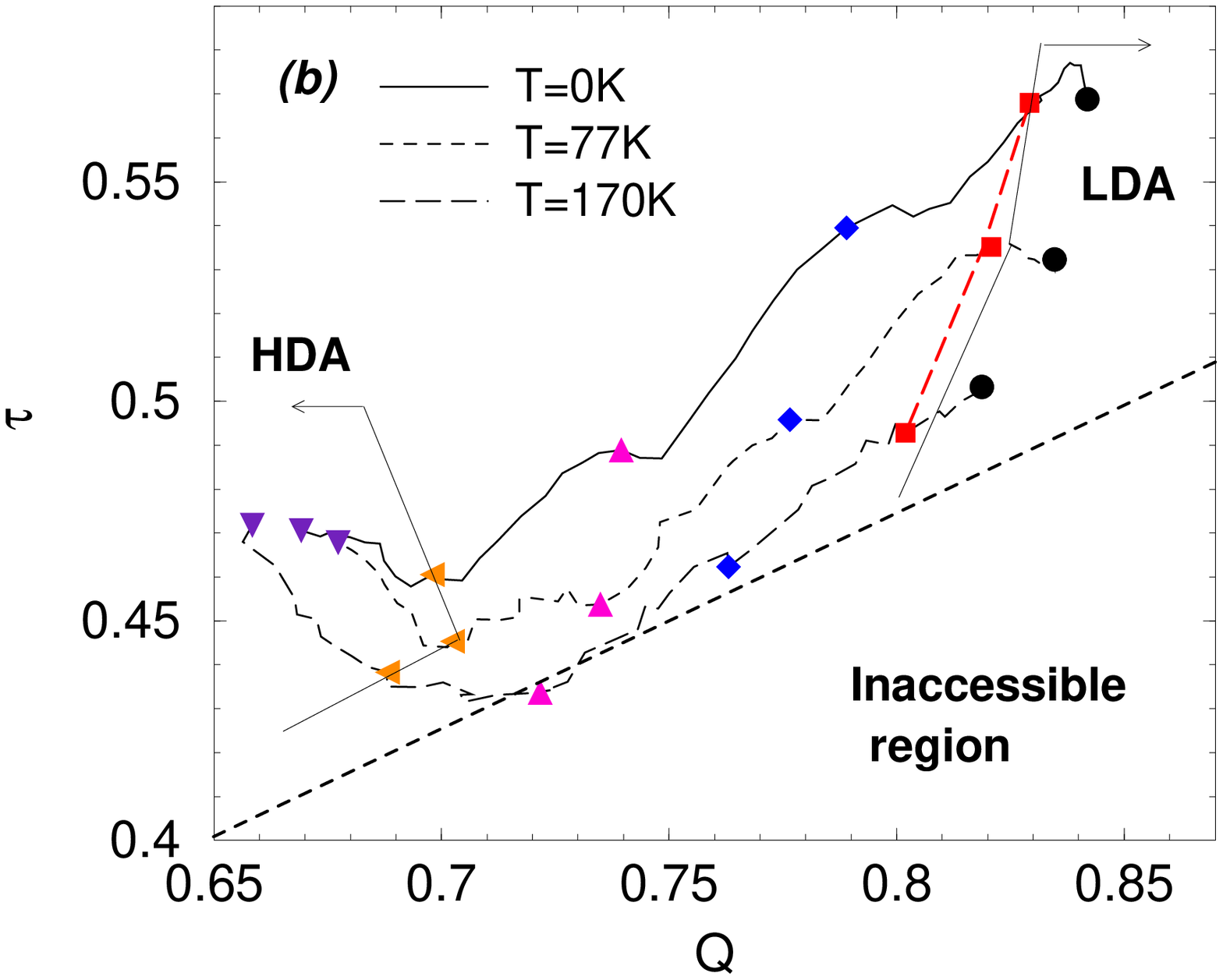}
}}
\vspace*{0.5cm}
\caption{(a) Evolution of pressure with density upon compression of 
low-density amorphous ice (LDA) to produce high-density amorphous ice (HDA). 
Compression is performed at $T=0,~77,$ and $170$~K, below the glass
transition temperature. We indicate approximately the location
corresponding to LDA and HDA.
 (b) Order parameters of the glasses sampled
along the isotherms shown in (a). The location in the order map corresponding
to LDA and HDA is also indicated.
Filled symbols correspond to the glasses obtained at
 $\rho=0.9$ up to $1.4$~g/cm$^3$ in steps of $0.1$~g/cm$^3$.}
\label{qt-ldahdavhda}
\end{figure}

Figure~\ref{qt-ldahdavhda}(a) shows the $\rho$-dependence of pressure $P$ 
during the LDA$\rightarrow$HDA transformation at the 
compression temperatures of $T=0,~77,$ and $170$~K, which lie below the glass
transition locus $T_g(P)$ for this system, as determined by the behavior 
of $C_P(T)$ upon heating (not shown).
 In experiments, LDA is obtained by isobaric
heating of HDA at $P=1$~atm. In computer simulations LDA has been obtained 
by isochoric cooling of a low-density liquid
\cite{poolepre,poole}. Following \cite{poolepre,poole}, we obtain the
starting LDA by {\it isochoric} fast-cooling of an equilibrium liquid at
$T=220$~K. 
The procedure followed to obtain LDA should not alter qualitatively our
results (see Appendix). In Fig.~\ref{qt-ldahdavhda}(a), LDA corresponds
 to the linear behavior of $P(\rho)$ at low-$\rho$.
 The transformation of LDA to HDA starts at the end
of this linear regime and is indicated on the left side of the figure.
 It is not so clear when the transformation to HDA is
complete, in particular because there are aging/annealing effects \cite{ourVHDAlong}. 
For simplicity, in the following analysis we assume that 
HDA corresponds to the glasses obtained by compression whose density exceeds 
$\rho_0 \equiv 1.3$~g/cm$^3$. The location of these glasses 
in the $P-\rho$ plane is indicated in Fig.~\ref{qt-ldahdavhda}(a).

Figure~\ref{qt-ldahdavhda}(b) shows the evolution in the order map 
of the glasses obtained during the LDA$\rightarrow$HDA transformation. 
The isotherms in Fig.~\ref{qt-ldahdavhda}(b) correspond to those
shown in Fig.~\ref{qt-ldahdavhda}(a). The symbols in each case denote the
densities $0.9,~1.0,~1.1,~1.2,~1.3,$ and $1.4$~g/cm$^3$. 
At the three temperatures studied, the orientational order
decreases monotonically upon compression,
 i.e., the glasses become less tetrahedral. However,
the translational order parameter 
behaves non-monotonically, exhibiting a minimum in the range from $1.25$ to
$1.3$~g/cm$^3$. An increase in translational order upon compression is the 
expected behavior for a condensed-phase system \cite{hards1,hards2,LennardJ}.
 In liquid water, this
behavior occurs only at high densities \cite{jeff}. It can be 
 seen from Fig.~\ref{qt-ldahdavhda}(b) that the same is true in glassy water.

It is interesting that LDA and
HDA are in well-defined and non-overlapping regions of the order map.
The difference in tetrahedrality is especially notable, with LDA-like
glasses having $Q$ values $\geq 0.8$ and HDA-like glasses having  
$Q$ values approximately $<0.7$.
We also note that the compressed glasses remain in the accessible region of
the order map.

\subsection{Annealed  Amorphous Ice}
\label{anneal}

\begin{figure}[p]
\narrowtext 

\centerline{
\hbox {
  \vspace*{0.5cm}  
  \epsfxsize=8cm
  \epsfbox{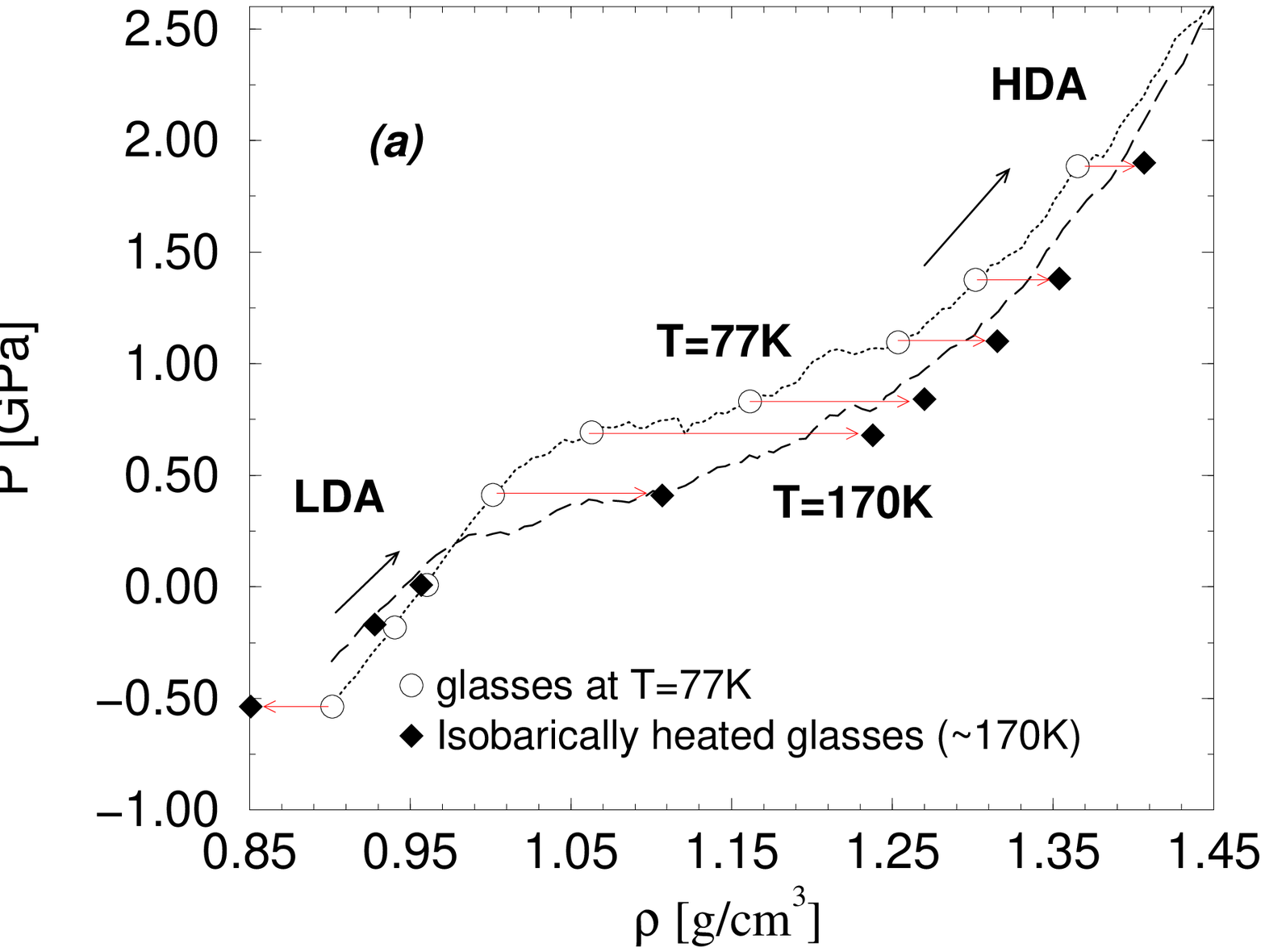}
}}
\centerline{
\hbox {
  \vspace*{0.5cm}  
  \epsfxsize=8cm
  \epsfbox{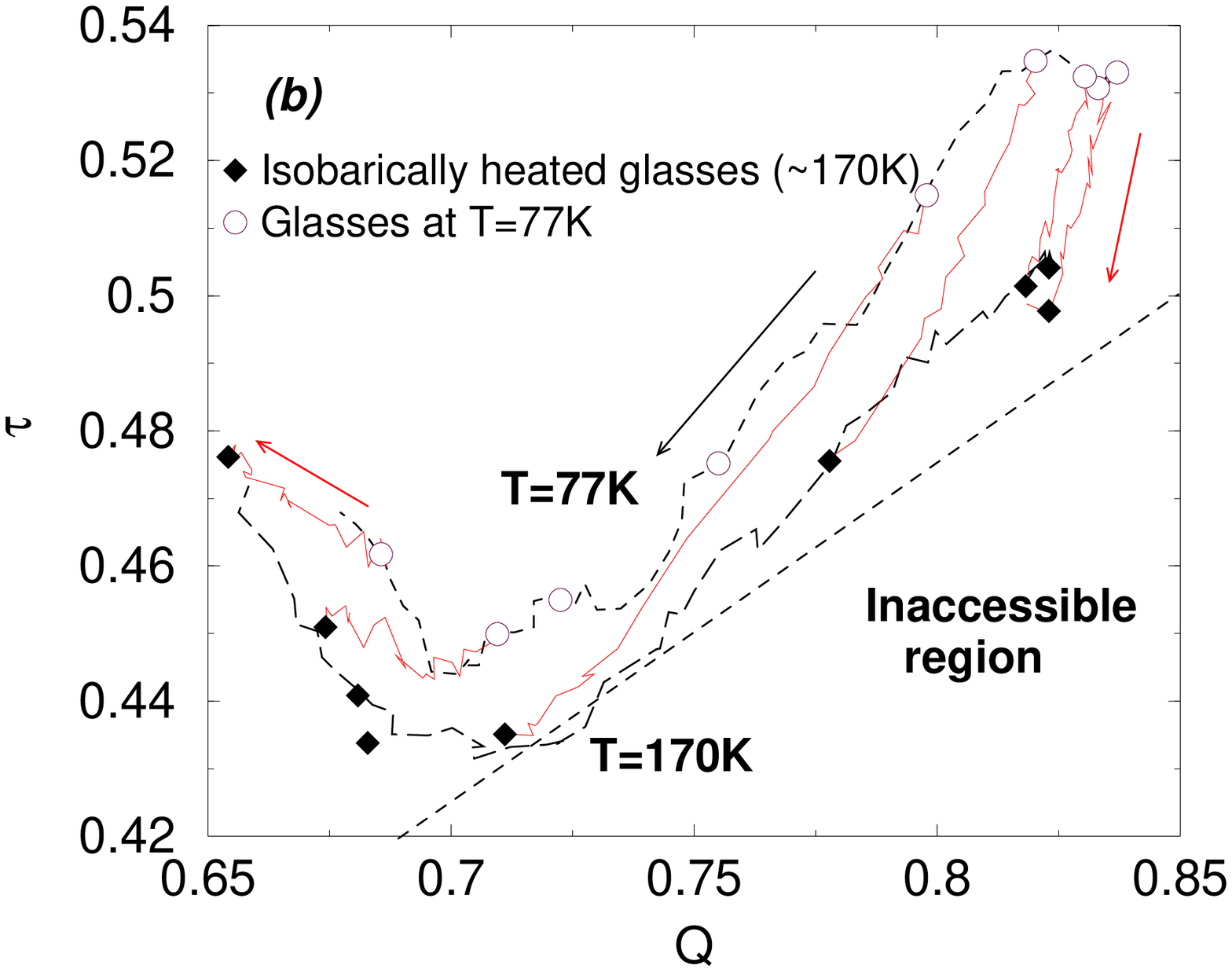}
}}
\vspace*{0.5cm}
\caption{(a) Evolution of pressure with density upon compression of 
low-density amorphous ice (LDA) to produce high-density amorphous ice (HDA). 
Compression is performed at $T=77,$ and $170$~K. Selected glasses
 at $T=77$~K 
 (open circles) are annealed/isobarically heated up
 to $T\approx 170$~K (filled diamonds); see
horizontal arrows. The annealed glasses in the $P-\rho$ plane approach or
cross the  $T\approx 170$~K-isotherm.
(b) Location in the order map of the glasses indicated in (a). Density
increases from right to left along both isotherms.}
\label{qt-ldahdavhda-170}
\end{figure}

Recently, it has been found that upon annealing or heating 
isobarically samples of HDA at
high-$P$, the glass becomes denser \cite{loerting}; the resulting glass is
called very-high-density glass (VHDA) or relaxed HDA (RHDA) 
\cite{ourVHDAlong,parrinello,ourVHDA}. VHDA/RHDA can also be
obtained in computer simulations using the SPC/E \cite{ourVHDA} and
the TIP4P models \cite{parrinello}.
In fact, it has been found \cite{ourVHDAlong} that annealing effects 
are present upon heating not only HDA, but also LDA and intermediate 
glasses in the LDA$\rightarrow$HDA transformation.
Figure~\ref{qt-ldahdavhda-170}(a) shows $P(\rho)$ during compression of LDA at
$T=77$~K and $T=170$~K. We also indicate the effect of isobarically heating 
 nine different glasses from $T=77$~K up to $T=170$~K
 \cite{ourVHDAlong} at pressures ranging from  $P=-0.55$ up to $1.9$~GPa (in
 the case of $P=-0.55$, the maximum $T$ reached is $T=160$~K). Upon heating, the
 density of the compressed glasses shifts from the ($T=77$~K)-isotherm 
(open circles) to the full diamonds located near the ($T=170$~K)-isotherm.

The changes in structural order that occur during annealing/isobaric heating 
of the nine glasses indicated in Fig.~\ref{qt-ldahdavhda-170}(a) are shown in
Fig.~\ref{qt-ldahdavhda-170}(b). 
The effect of annealing LDA is mainly to reduce
the translational order while only slightly reducing the orientational order.
This is clear from the isobaric heating of LDA, at
$P=-0.55,~-0.17,$ and $0.01$~GPa [trajectories starting from the first three
circles (high-$Q$) of the $77$~K-isotherm in Fig.~\ref{qt-ldahdavhda-170}(b)].
 However, as $P$ is increased approaching the LDA$\rightarrow$HDA
 transformation region, the tetrahedral order becomes more sensitive
 to annealing. 
This is clear from the heating at $P=0.41$~GPa, which corresponds 
 to a glass located at the end of the LDA region in 
Fig.~\ref{qt-ldahdavhda-170}(a) [trajectory starting from the fourth circle in
Fig.~\ref{qt-ldahdavhda-170}(b)].
Although the order
parameters indicate a considerable change in structure,
in the case of the isobaric heating at
$P=0.01$~GPa, the density of the glass barely changes.

In contrast to the behavior observed upon isobaric heating of LDA, 
annealing HDA mainly reduces
the orientational order, while small changes in $\tau$ are found to occur.
This is clear from the isobaric heatings at $P=1.38$ and $1.9$~GPa.
 [trajectories starting from the last two circles (low-$Q$) of the
 $77$~K-isotherm in Fig.~\ref{qt-ldahdavhda-170}(b)].
 We note that in the case of the isobaric heating at $P=1.9$~GPa, there is 
actually an increment in translational order, the opposite effect to that
 observed upon heating LDA.
In the case of isobaric heating at intermediate pressures (approximately
$0.41$~GPa~$< P <1.38$~GPa), both the translational and orientational order 
parameters decrease appreciably.

We stress that upon annealing HDA ($P=1.38$ and $1.9$~GPa), i.e. during the 
 HDA$\rightarrow$VHDA transformation, 
the location of the glass in the order map shifts to a low-$Q$ subregion
 already included in the HDA domain. Similarly, 
annealing LDA ($P=-0.55,~-0.17,$ and $0.01$~GPa)
 shifts the location of the glass in the order map to a low-$\tau$ subregion 
already included in the LDA domain.
 This does not apply to glasses of intermediate densities 
($0.98$~g/cm$^3$ $< \rho < 1.25$~g/cm$^3$, approximately) which cannot be
characterized as LDA or HDA before annealing. These systems start from,
and evolve toward, an intermediate-$Q$ region in the order map
($\approx 0.7 \leq Q < 0.8$).  

A salient feature of Fig.~\ref{qt-ldahdavhda-170}(a)
and~\ref{qt-ldahdavhda-170}(b) is the similarity, both in density and in
structural order, between isobarically annealed glasses at $T=170$~K and
isothermally ($T=170$~K) compressed glasses at the same density.

\section{Comparison of Structural Order in Isothermally Compressed,
  Isobarically Annealed, and Isochorically Cooled Glasses}
\label{comparisonGlasses}

\subsection{Isochorically Cooled Glasses and the Low-density and High-density Amorphous Ices}

\begin{figure}[p]
\narrowtext 

\centerline{
\hbox {
  \vspace*{0.5cm}  
  \epsfxsize=8cm
  \epsfbox{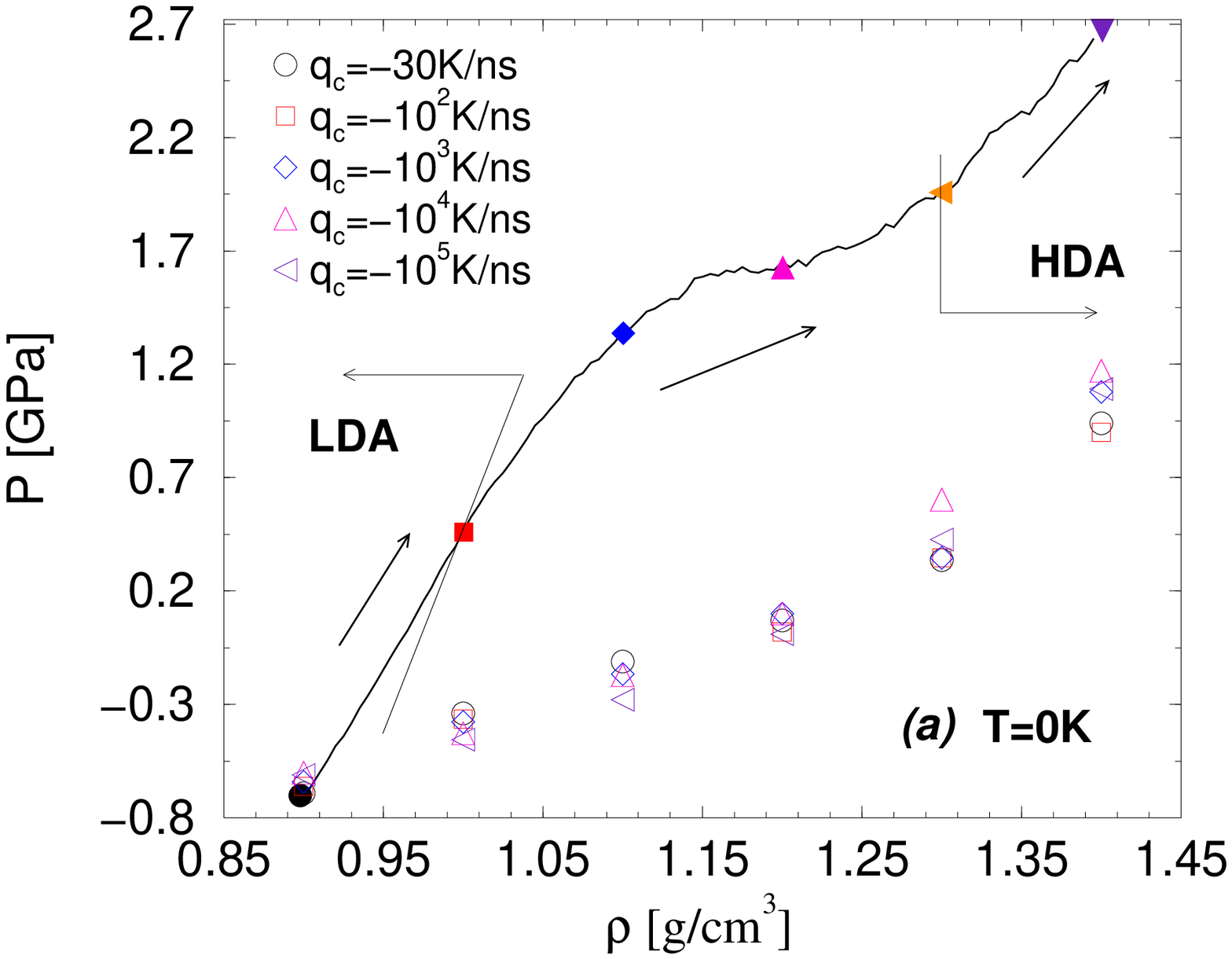}
}}
\centerline{
\hbox {
  \vspace*{0.5cm}  
  \epsfxsize=8cm
  \epsfbox{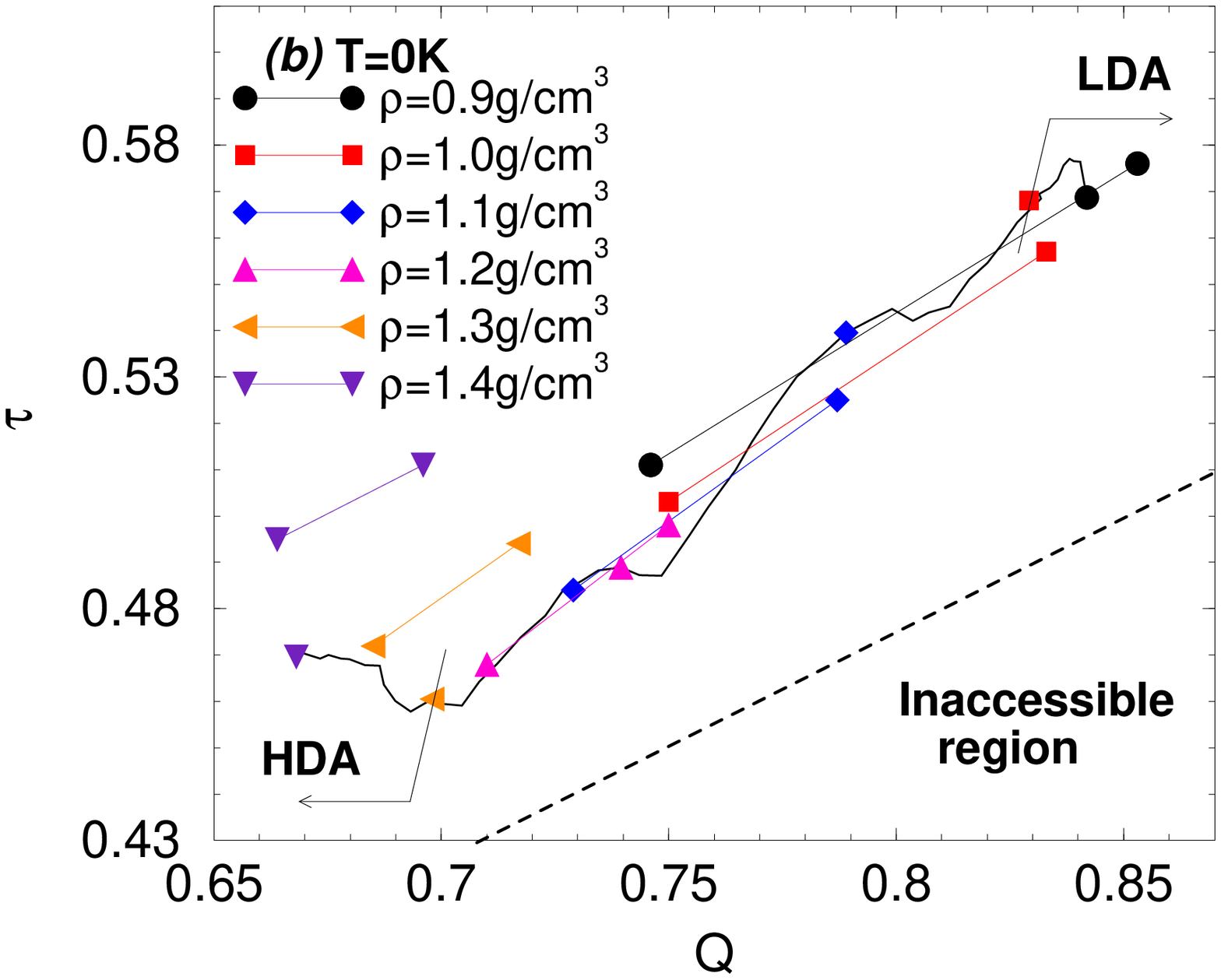}
}}
\vspace*{0.5cm}
\caption{(a) Comparison of the location in the $P-\rho$ plane of the glasses
  at $T=0$~K obtained by isothermal compression of LDA and by isochorically
  cooling at different densities and cooling rates. At $T=0$~K no
  annealing is possible.
 Open symbols represent different cooling
  rates while solid symbols on the $T=0$~K-isotherm correspond to glasses from
  $\rho= 0.9$~g/cm$^3$ to $\rho=1.4$~g/cm$^3$ in steps of $0.1$~g/cm$^3$. 
(b) Location in the order map of the glasses indicated in (a).
Glasses formed  by isochoric cooling lie on the straight lines connecting
fast- (low-$Q$ end) and slow-cooled (high-$Q$ end) extremes at each density.
We also show  the  location of the glasses along the
  $T=0$~K-isotherm indicated in (a). 
}
\label{hgw-ldahda-0K}
\end{figure}

We first compare structural order in LDA and HDA with that in
isochorically-quenched glasses (Section~\ref{cooling}) at the same density.
To avoid aging effects, we compare the glasses
obtained upon compression of LDA at $T=0$~K to the isochorically-cooled
glasses obtained also at $T=0$~K.
The location of these
 isochorically-quenched glasses in the $P-\rho$ plane is indicated by open
symbols in Fig.~\ref{hgw-ldahda-0K}(a). 
Full symbols denote the glasses from the 
LDA$\rightarrow$HDA isothermal compression at $T=0$~K.
 Clearly, glasses obtained upon
compression have very different pressures than isochorically-cooled
 glasses of the same density over the range of compression
and cooling rates explored here. In fact, we find that 
the glasses obtained by 
 compression of LDA (at the present compression rate) at a given $\rho$ 
are less stable
 than those obtained by slow cooling at the same density.
 In other words, $e_{IS}$ for the
 compressed glasses is higher than for the
 isochorically cooled glasses at $q_c=-30$~K/ns.

These glasses are also different structurally. In
Fig.~\ref{hgw-ldahda-0K}(b) we compare the evolution of the isothermal
LDA$\rightarrow$HDA compression at $T=0$~K in the $(Q, \tau)$ plane with the
corresponding location of isochorically quenched glasses. The 
LDA$\rightarrow$HDA compression is represented by the irregular curve. For
the isochorically-quenched glasses we show the $(Q, \tau)$ coordinates of the
slowest- and fastest-cooled glasses ($q_c=-30$~K/ns, high-$Q$ end; 
$q_c=-10^5$~K/ns, low-$Q$ end, respectively), joined by a straight line,
along which lie the glasses cooled at intermediate rates [see, e.g., 
Fig.~\ref{qt-plane}(a)]. The symbols denote different densities, as indicated in
the figure.

Among the low-$\rho$ isochorically-cooled glasses ($\rho=0.9$~and
$\rho=1.0$~g/cm$^3$), only those cooled at the slowest rates ($q_c=-30$
and $q_c=-10^2$~K/ns) fall inside the LDA region. For
$|q_c|>10^2$~K/ns we find that the structure corresponds to
 glasses lying between LDA and HDA in the $(Q, \tau)$ plane.
 As previously shown \cite{ourPRE-RC},
glasses obtained at rates $|q_c|>10^2$~K/ns are highly metastable and as soon
as they are heated, they relax toward those glasses obtained by slower
cooling rates.

At high densities, $\rho>1.2$~g/cm$^3$, isochorically-quenched 
 glasses do not fall into the HDA region in the order map,
 as one would expect based on their density. 
Even the slow-cooled glasses are located in the region between that of LDA
and HDA. 
This again highlights the fact
 that the glasses obtained by compression in the LDA$\rightarrow$HDA
transformation are structurally quite distinct from
 those obtained by isochoric
cooling (at least with the cooling and compression rates 
accessible with the present computer times). 

In fact, we have found (results are not shown here)
 that $e_{IS}$, ${\cal S}_{IS}$, and $P_{IS}$ of the glasses 
obtained upon isochoric cooling at a given density 
and those obtained upon compression of LDA at the same density 
can be very different and are very sensitive to the cooling rate. 
A similar situation seems to happen when comparing glasses obtained upon
isobaric cooling with those from the LDA$\rightarrow$HDA transformation.
For example, VHDA/RHDA can be obtained upon 
isobaric cooling only upon fast-cooling while HDA is not accessible upon
isobaric cooling of the liquid \cite{ourVHDA}.

\subsection{Isochorically Cooled Glasses and the  Annealed Amorphous Ices}

\begin{figure}[p]
\narrowtext 

\centerline{
\hbox {
  \vspace*{0.5cm}  
  \epsfxsize=8cm
  \epsfbox{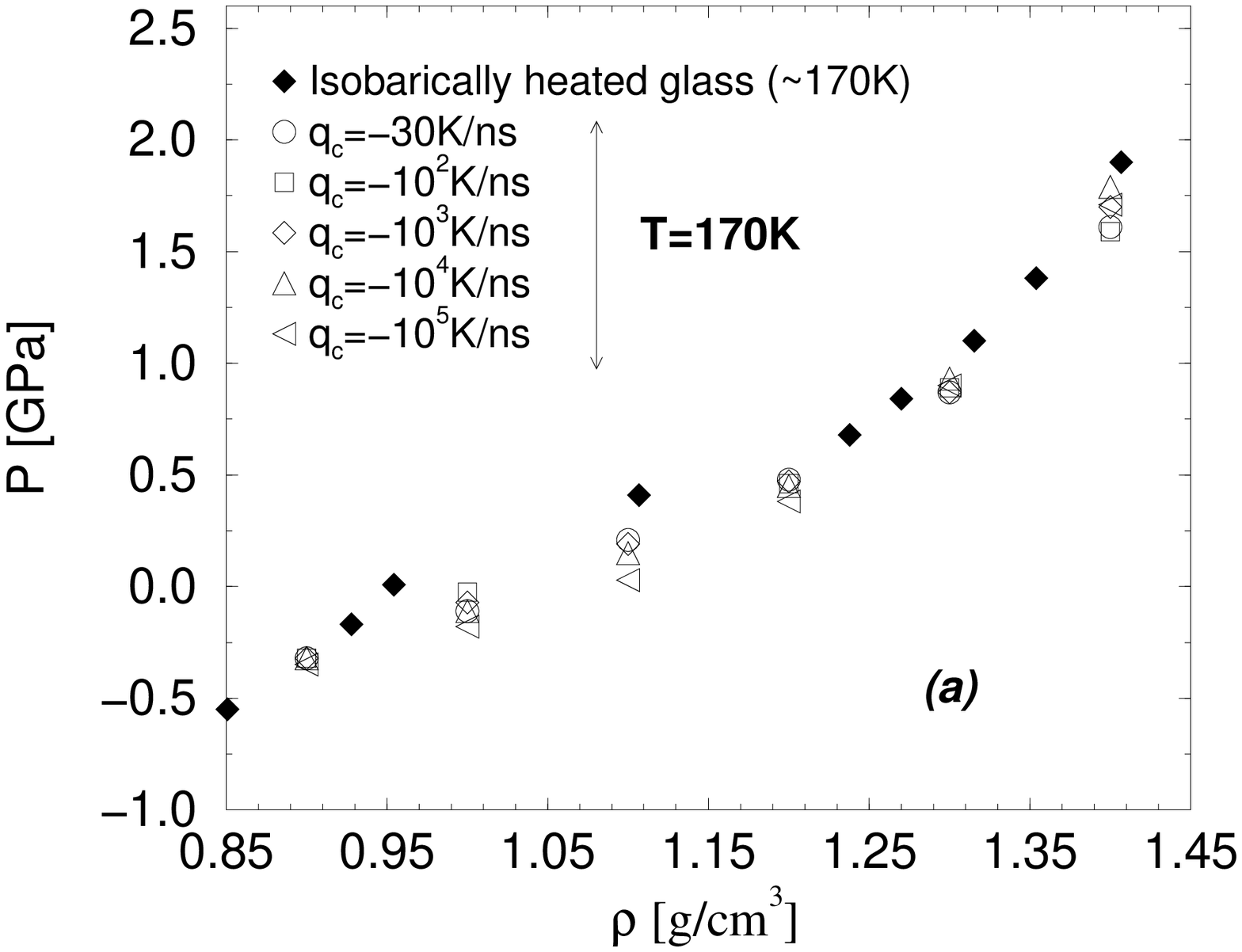}
}}
\centerline{
\hbox {
  \vspace*{0.5cm}  
  \epsfxsize=8cm
  \epsfbox{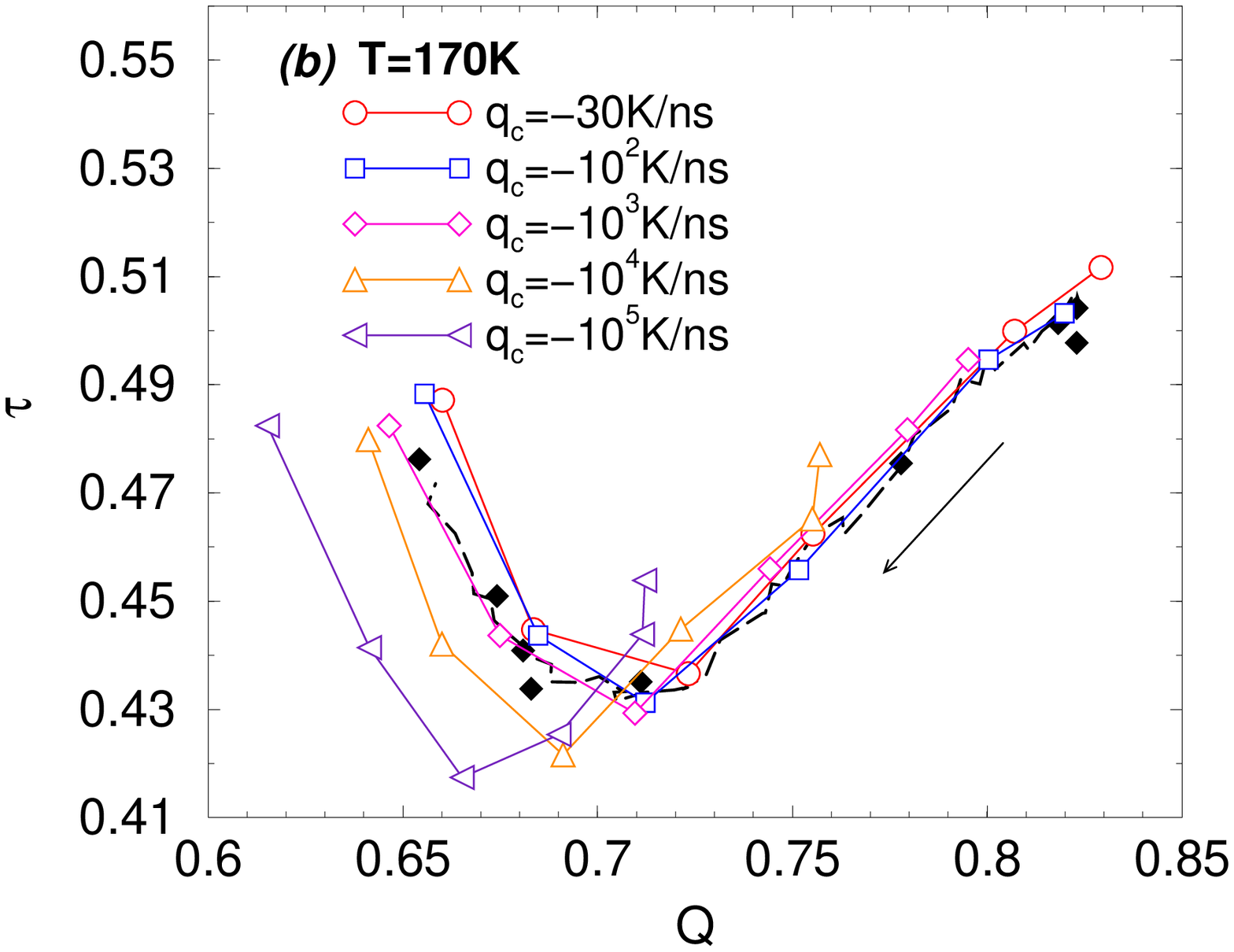}
}}
\vspace*{0.5cm}
\caption{(a) Comparison of the location in the $P-\rho$ plane of the glasses
  at $T=170$~K obtained by 
 isochoric cooling at different densities and cooling rates (open
 circles), and by annealing/isobarically heating of glasses obtained by
 compression of LDA at $T=77$~K (solid diamonds) [see
 Fig.~\ref{qt-ldahdavhda-170}(a)]. 
(b) Location in the order map of the glasses indicated in (a). Lines
connecting open symbols correspond to glasses obtained at a given cooling rate
for different densities [density increases
from $0.9$~g/cm$^3$ (high-$Q$ values) to $1.4$~g/cm$^3$ (low-$Q$ values), in
steps of $0.1$~g/cm$^3$].
For comparison, we also show the location of the glasses obtained by 
isothermal compression ($T=170$~K) of LDA (long-dashed line).
We note that all glasses in (b) are at $T=170$~K
 and, therefore, they have the same vibrational contributions.  
}
\label{hgw-ldahda-170K}
\end{figure}

Figure~\ref{hgw-ldahda-170K}(a) shows the location in the
$P$-$\rho$ plane of the annealed/isobarically heated
glasses indicated in Fig.~\ref{qt-ldahdavhda-170}(a) (diamond symbols) and 
 of the glasses cooled isochorically down to $T=170$~K
 at different cooling rates (open symbols).
Comparison of Fig.~\ref{hgw-ldahda-0K}(a) and~\ref{hgw-ldahda-170K}(a) shows
that at this relatively high temperature, equation-of-state differences between glasses
formed by different routes, though still evident, are much less pronounced.

The location in the order map of the glasses indicated in
 Fig.~\ref{hgw-ldahda-170K}(a) is shown in Fig.~\ref{hgw-ldahda-170K}(b).
For comparison, we also
include the location of the glasses obtained by isothermal 
compression at $T=170$~K (long-dashed line). The annealed glasses are
structurally very similar to those obtained upon isothermal compression,
 i.e., these glasses fall on the $T=170$~K-compression curve in   
  Fig.~\ref{hgw-ldahda-170K}(b). 

We also show in the same figure the ``isotherms'' connecting glasses formed
by isochoric cooling at the same rate. Symbols identify the quench rate, and
along each ``isotherm'' the density increases monotonically from
$0.9$~g/cm$^3$ (high-$Q$ end) to $1.4$~g/cm$^3$ (low-$Q$ end) in
$0.1$~g/cm$^3$ increments.
Note that neighboring points along an ``isotherm'' are not connected by a
continuity of states, but were instead attained by isochoric quenches at
different densities. In all these cases, $Q$
decreases monotonically upon compression, while $\tau$ decreases initially 
 up to $\rho \approx 1.0$~g/cm$^3$ and
 increases thereafter. With exception of the isotherms
corresponding to the fastest cooling rates
($q_c=-10^4,~-10^5$~K/ns), the low-density portions 
($\rho \leq 1.25$~g/cm$^3$) of all the curves collapse onto
 a common isotherm.
We thus find that glasses with $0.9$~g/cm$^3$$\leq \rho \leq 1.3$~g/cm$^3$,
when sufficiently hot, share with the cold equilibrium liquid the remarkable 
feature \cite{jeff} of having only one independent order parameter, it being
impossible to change one (e.g., $Q$) without changing the other.

We note, finally, that at high enough densities, appreciable structural
differences between glasses formed by different routes are evident even when
their thermal properties are very similar.
For example, all the isochorically cooled glasses at
$\rho=1.3$~g/cm$^3$ are at almost the same $P$ [see
Fig.~\ref{hgw-ldahda-170K}(a)]. However,    
Fig.~\ref{hgw-ldahda-170K}(b) shows that these glasses have very 
different order parameters (next-to-last open symbols on each isotherm, on
the low-$Q$ end).

\section{Conclusions}
\label{conclusions}

In this work we have used order metrics $Q$ and $\tau$ and the order map concept
\cite{hards1,hards2,LennardJ,jeff}, to quantitatively characterize structural
order in glassy states of the SPC/E model of water. Our main findings are
recapitulated below. 

At a given density, glasses formed by isochoric cooling down to 
 $T=0$~K define a line in the order map. This
  line shifts non-monotonically across the order map upon changing the 
 density, and corresponds also to the location in the $(Q, \tau)$ 
plane of the IS sampled by the system both {\it in and out of equilibrium}.
The finding that the order parameters of the IS fall (for a given density) 
on a line in the order map implies that $Q$ and $\tau$ are not
independent. In fact, we find that both $Q$ and $\tau$ 
 are linear functions of the depth of the IS, i.e. $e_{IS}$.

We have also investigated in detail the structural order of the glasses formed
upon isothermal compression across the LDA$\rightarrow$HDA transformation.
At the three temperatures studied ($0,~77$ and $170$~K), the orientational order decreases
monotonically upon compression.
However, the translational order decreases until $\rho \approx
1.3$~g/cm$^3$ (approximately, the HDA formation density); further compression
increases the translational order.
We were able to identify two non-overlaping and
well-defined regions: a high-Q and high-$\tau$ region corresponding to LDA, and 
a low-Q and low- or intermediate-$\tau$ region, corresponding to HDA.

Annealing (i.e., isobaric heating) of compressed glasses is 
important experimentally: it has been found that
it can change dramatically their properties \cite{mishima96,loerting,finneyHDAVHDA}.
 Accordingly, we also have studied the evolution of structural order in
 isobarically-annealed glasses over  a wide range
of $P$. We find that annealing LDA not only decreases the density but
also causes a significant decrease in translational order,
 while the orientational order barely
changes. Annealing HDA (this corresponds to the HDA$\rightarrow$VHDA/RHDA
transformation \cite{parrinello,ourVHDA,loerting}), reduces the orientational
order appreciably, and causes a slight increase in translational order.
While structural changes upon annealing are significant, they do not cause
LDA or HDA to move outside of their respective regions in the order map
(defined {\it before} annealing).

We compared the structure of glasses formed by different routes in order to
 test whether we can relate the LDA/HDA glasses with those obtained by isochoric
cooling at the same density 
(at least, with the present cooling/compression rates available in
computer simulations).
The structure of the glasses formed by different routes can be compared by
traditional methods, e.g., the oxygen-oxygen radial distribution function
$g_{OO}(r)$, or, alternatively, by the order map characterization utilized here.
When comparing the structure of  isochorically cooled, and compressed
glasses at same $T$ and $\rho$, we find that the location in the order map is very
 sensitive to structural changes.
Among isobarically-cooled glasses at $T=0$~K and 
$\rho=0.9$ or $1.0$~g/cm$^3$, only those cooled at the slowest rates
($q_c=-30$ and $-10^2$~K/ns) fall into the LDA region in the order map.
In the case of the isochorically-cooled glasses at
$T=0$~K and $\rho=1.3$ or $1.4$~g/cm$^3$, they fall outside of the HDA region
in the order map. Therefore, isochorically-cooled glasses in the appropriate density range
 cannot always be classified as LDA or HDA.

In contrast to the pronounced path-dependence of the structure of `cold'
glasses, we find much weaker history dependence in the properties of glasses
 formed after isobaric annealing at
$T=170$~K, when compared with those attained 
 by isochoric cooling glasses down to the same $T$.
 Both the annealed and slow-cooled glasses are structurally
 very similar, and they are
located in the order map on the same isotherm as 
compressed glasses at $T=170$~K. 
We also find that glasses formed by isochoric cooling to $T=170$~K collapse
onto a common `master' isotherm in the order map regardless of their density
as long as $|q_c|<10^4$~K/ns.

As a final observation, we note that all the glasses obtained in this work, under any
condition and procedure fall in the accessible
 region of the order map, as found in \cite{jeff} for liquid water.
 This suggests that the inaccessible region (at least for systems lacking  long-range order)
is a consequence of constrains  inherent to the molecular interactions and
 is not dependent on whether the system is a liquid or a glass.

In this work we have applied the order map methodology to the investigation
of structural order in water glasses. We find that location in the order map
is a sensitive descriptor of structure.
 We further find that at low enough temperature, glasses obtained by
isochoric cooling are very different from those formed by isothermal
compression, even when compared at the same density and temperature.
We were able to identify distinct regions of the order map corresponding to
LDA ($Q \geq 0.8, \tau > 0.46$), and HDA ($Q \leq 0.7, 0.43 < \tau < 0.5$),
and to follow the evolution of structural order upon isobaric
annealing. These results demonstrate the usefulness of an order metric approach
in the quest for a more precise characterization of structural order in
amorphous systems in general and water glasses in particular.

\section*{Acknowledgments}
We thank J.R. Errington for providing the numerical data
 shown in Fig.~5 of Ref.~\cite{jeff}.  
PGD and HES gratefully acknowledge the support of NSF through Collaborative
Research in Chemistry Grant No. CHE 0404699. FS acknowledges the support of
 MIUR-Firb 2002.

\section*{Appendix}
\label{append}

To test how sensitive the simulation results are to the method of preparation of LDA, 
we compare the compression of LDA prepared by {\it
  isochoric} cooling of the equilibrium liquid at $\rho=0.9$~g/cm$^3$,
 with that of LDA prepared by {\it isobaric} cooling
 of equilibrium liquid at $P=0.1$~GPa. In fact, this last procedure
 is the recipe followed in experiments to obtain
hyperquenched-glassy water, HGW \cite{HGW}. X-ray and
neutron diffraction measurements suggest that LDA is structurally identical
 to HGW \cite{bellisent} and, although small differences have been
 found\cite{waterAB,klugPRL,TseKlug}, the common view at present is that HGW
 and LDA are the same material
 \cite{PabloReview,geneNature,genePhysToday}.

\begin{figure}[p]
\narrowtext 

\centerline{
\hbox {
  \vspace*{0.5cm}  
  \epsfxsize=8cm
  \epsfbox{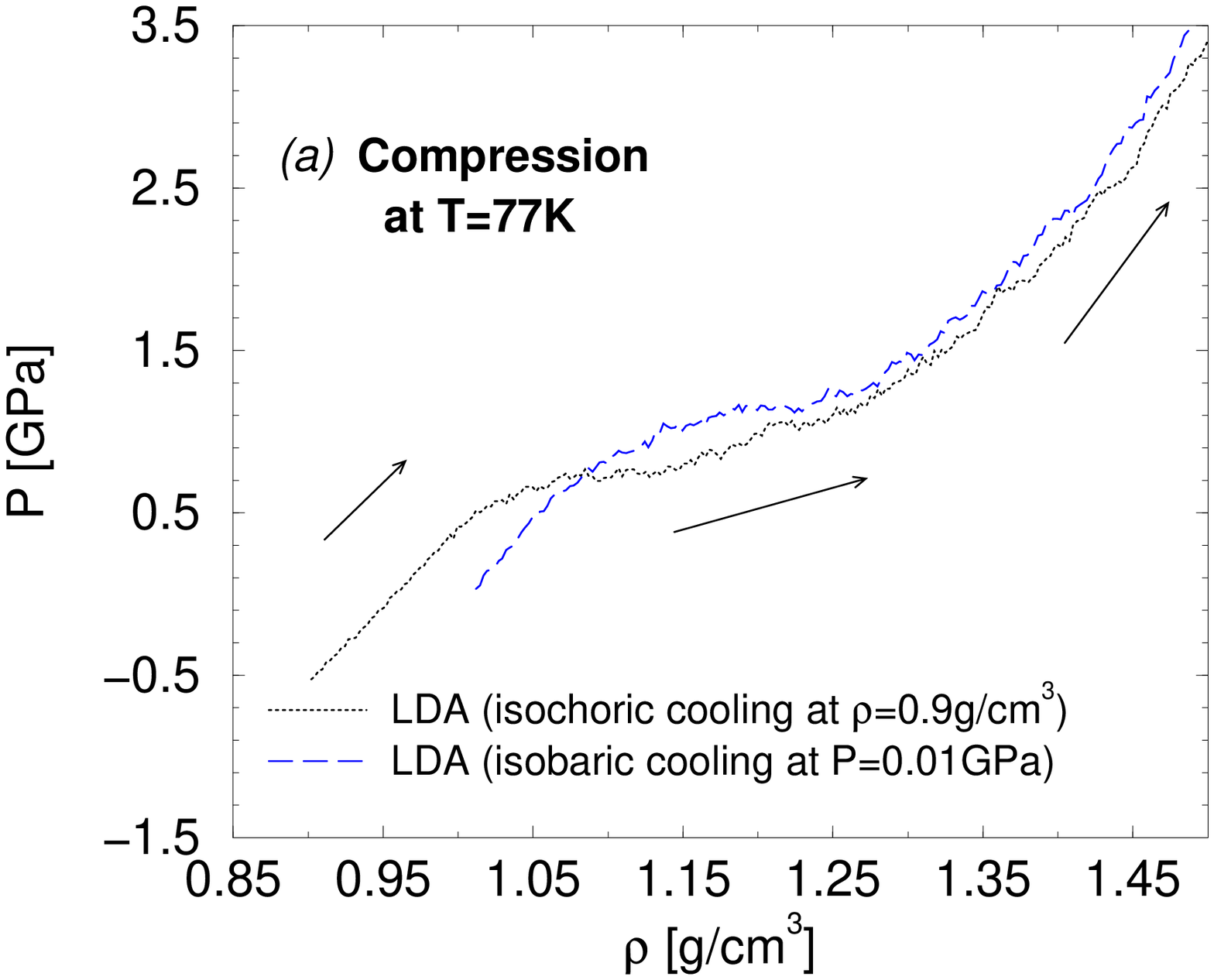}
}}
\centerline{
\hbox {
  \vspace*{0.5cm}  
  \epsfxsize=8cm
  \epsfbox{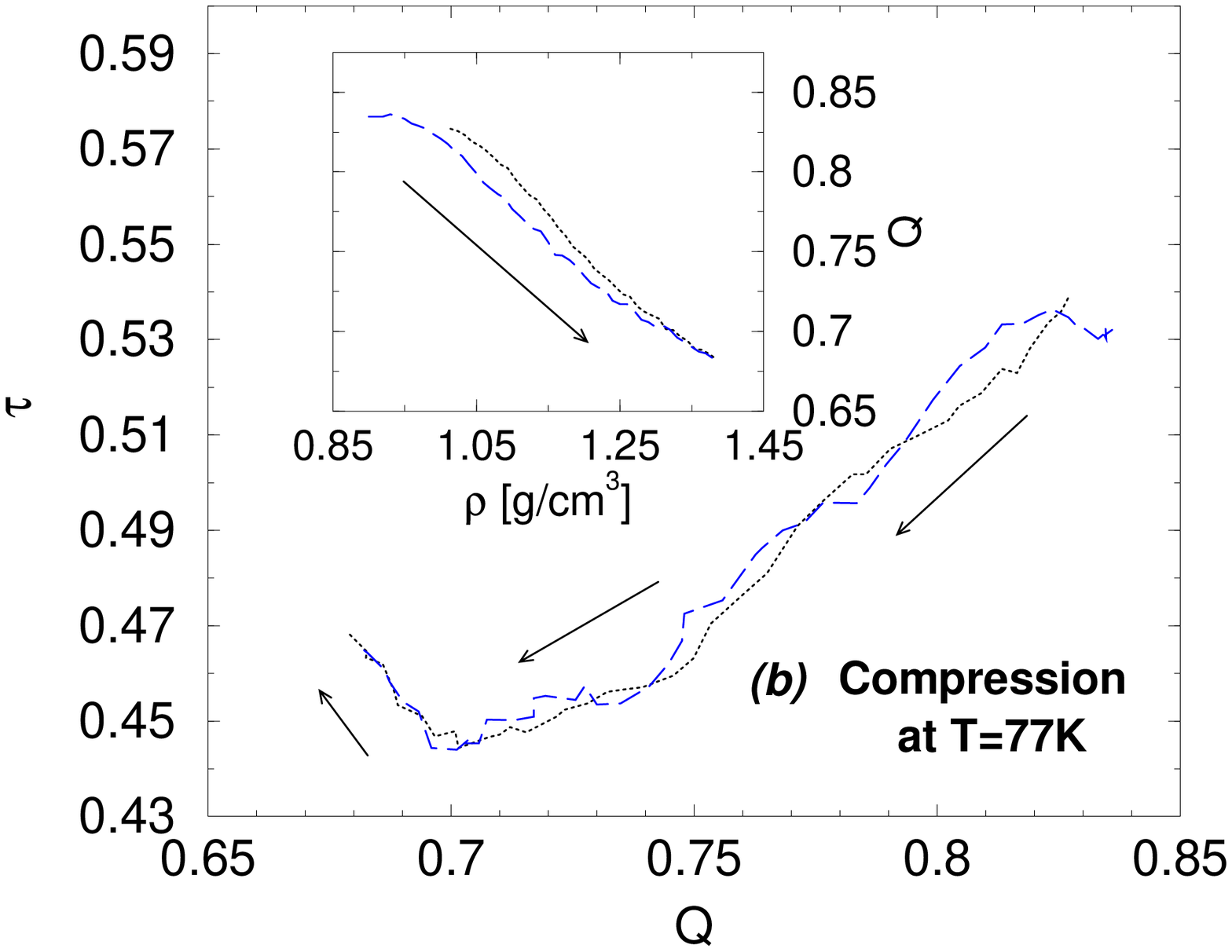}
}}
\vspace*{0.5cm}
\caption{Comparison of the $\rho$-dependence upon compression for two samples
of LDA prepared in a different way: LDA obtained by isochoric fast-cooling of 
equilibrium liquid at $T=220$~K and $\rho=0.9$~g/cm$^3$, and LDA obtained by
isobaric slow-cooling at $P=0.01$~GPa. (b) Comparison of the order
parameters in the glasses sampled upon compression both samples of LDA shown
in (a). Results are qualitatively similar in both cases suggesting that the
preparation of LDA does not alter our conclusions.}
\label{lda-hgw}
\end{figure}

Figure~\ref{lda-hgw}(a) shows the $\rho$-dependence of $P$ when compressing
both samples of LDA at $T=77$~K. Both $P(\rho)$ curves behave in a qualitatively similar
way. It can be seen that the two compression curves are similar, especially
at high enough densities. 
Equation-of-state properties reveal some sensitivity to sample history at
low densities.

Figure~\ref{lda-hgw}(b) shows the corresponding trace in the order map
of the two compressions. We observe 
that both samples of LDA exhibit a very similar structural evolution.
Therefore, Figure~\ref{lda-hgw} suggests that our results do not depend 
strongly on the way in which LDA is prepared.

\newpage

\newpage

\newpage

\newpage

\newpage

\newpage

\newpage

\newpage

\newpage

\begin{references}

\bibitem{kittel}
C. Kittel, {\it Introduction to Solid State Physics}, 7th ed. (Wiley, New
York, 1996).

\bibitem{zallen}
R. Zallen, {\it The Physics of Amorphous Solids}, (Wiley, New
York, 1983).

\bibitem{torquato}
S. Torquato, {\it Random Heterogeneous Materials: Microstructures and
Macroscopic Properties} (Springer-Verlag, New York, 2002).

\bibitem{hards1}
S. Torquato, T.M. Truskett, and P.G. Debenedetti,
Phys. Rev. Lett. {\bf 84}, 2064 (2000).

\bibitem{hards2}
T.M. Truskett, S. Torquato, and P.G. Debenedetti,
Phys. Rev. E {\bf 62}, 993 (2000).

\bibitem{LennardJ}
J. R. Errington, P.G. Debenedetti, and S. Torquato,
J. Chem. Phys. {\bf 118}, 2256 (2003).

\bibitem{nelson}
P.J. Steinhardt, D.R. Nelson, and M. Ronchetti, 
Phys. Rev. B {\bf 28}, 784 (1983).

\bibitem{jeff}
J.R. Errington and P.G. Debenedetti, Nature {\bf 409}, 318 (2001).

\bibitem{6to9}
G. Ruocco, M. Sampoli, A. Torcini, and R. Valluari,
J. Chem. Phys. {\bf 99}, 8095 (1993); F.X. Prielmeier, E.W. Lang, 
R.J. Speedy, and H.-D. L\"{u}deman,
Phys. Rev. Lett. {\bf 59}, 1128 (1987); C.A. Angell, E.D. Finch, L.A. Woolf,
and P. Bach, J. Chem. Phys. {\bf 65}, 3063 (1976).

\bibitem{francislong}
F. W. Starr, F. Sciortino, and H.E. Stanley, 
Phys. Rev. E {\bf 60}, 6757 (1999).

\bibitem{AntonioNature}
A. Scala, F.W. Starr, E. La Nave, F. Sciortino, and H.E. Stanley,
Nature {\bf 406}, 166 (2000).

\bibitem{pabloBook}
P.G. Debenedetti, {\it Metastable Liquids, Concepts and Principles},
(Princeton University Press, Princeton, 1996).

\bibitem{qt-silica}
M.S. Shell, P.G. Debenedetti, and A.Z. Panagiotopoulos,
Phys. Rev. E {\bf 66}, 011202 (2002).

\bibitem{PabloReview}
P.G. Debenedetti, J. Phys.: Condens. Matter {\bf 15}, R1669 (2003).

\bibitem{AngellReview}
C.A. Angell, Ann. Rev. Chem. {\bf 55}, 559 (2004). 

\bibitem{geneNature} O. Mishima and H. E. Stanley, Nature {\bf 396}, 329
(1998).

\bibitem{mishima84}
O. Mishima and L.D. Calvert, and E. Whalley, Nature {\bf 310}, 393 (1984).

\bibitem{floriano} M. A. Floriano {\it et al.}, 
Y. P. Handa, D. D. Klug and E. Whalley,
J. Chem. Phys. {\bf 91}, 7187 (1989). 

\bibitem{joharyIc}
G.P. Johari, A. Hallbrucker, and Mayer, J. Phys. Chem. {\bf 94},
1212 (1990). 

\bibitem{mishimaNature85} O. Mishima, L. D. Calvert and E. Whalley,
Nature {\bf 314}, 76 (1985). 

\bibitem{mishimaVisual} O. Mishima, K. Takemura and K. Aoki, Science
{\bf 254}, 406 (1991). 

\bibitem{mishimaJCP}
 O. Mishima, J. Chem. Phys. {\bf 100}, 5910 (1993). 

\bibitem{mishimaLL} O. Mishima and H. E. Stanley, Nature {\bf 392}, 164
(1998).

\bibitem{newKlotz}
S. Klotz {\it et al.}, 
Phys. Rev. Lett. {\bf 94}, 025506 (2005).

\bibitem{mishima96}
O. Mishima, Nature {\bf 384}, 546 (1996).

\bibitem{ourVHDAlong}
N. Giovambattista, H.E. Stanley, and F. Sciortino, submitted; 
cond-mat/0502531.

\bibitem{johary2004}
G. P. Johari and O. Andersson, J. Chem. Phys. {\bf 120}, 6207 (2004).   

\bibitem{science} C. A. Tulk, 
C. J. Benmore, J. Urquidi, D. D. Klug, J. Neuefeind, B. Tomberli 
and P. A. Egelstaff, Science {\bf 297}, 1320 (2002).

\bibitem{parrinello}
R. Marto\v{n}\'{a}k, D. Donadio, and M. Parrinello, 
Phys. Rev. Lett. {\bf 92}, 225702 (2004).

\bibitem{ourVHDA}
N. Giovambattista, H.E. Stanley, and F. Sciortino, Phys. Rev. Lett. (to be
published); cond-mat/0403365.

\bibitem{french}
B. Guillot and Y. Guissani, J. Chem. Phys. {\bf 119}, 11740 (2003).

\bibitem{berensen}
H.J.C. Berendsen, J.R. Grigera and T.P. Stroatsma,
J. Phys. Chem. {\bf 91}, 6269 (1987). 

\bibitem{pooleSPC/E} 
S. Harrington, P. H. Poole, F. Sciortino, and H. E. Stanley,
J. Chem. Phys. {\bf 107}, 7443 (1997). 

\bibitem{francescoSPCE2}
 F. Sciortino, L. Fabbian, S.-H. Chen, and P. Tartaglia,
 Phys. Rev. E {\bf 56}, 5397 (1997).

\bibitem{ourSHD}
N. Giovambattista, S.V. Buldyrev, F.W. Starr, and H.E. Stanley,
Phys. Rev. Lett. {\bf 90}, 085506 (2003).

\bibitem{francescoSPCE1}
 F. Sciortino, P. Gallo, P. Tartaglia, and S.-H. Chen,
 Phys. Rev. E {\bf 54}, 6331 (1996).

\bibitem{ourLDAHDA} 
N. Giovambattista, H.E. Stanley, and F. Sciortino, 
Phys. Rev. Lett. {\bf 91}, 115504 (2003). 

\bibitem{berendsenThermo}
H.J.C. Berendsen, J.P.M. Postma, W.F. van Gunsteren, A. DiNola, and J.R. Haak,
J. Phys. Chem. {\bf 81}, 3684 (1984). 

\bibitem{ourSHADOW}
N. Giovambattista, C.A. Angell, F. Sciortino, and H.E. Stanley, 
Phys. Rev. Lett {\bf 93}, 047801 (2004). 

\bibitem{ourPRE-RC}
N. Giovambattista, H.E. Stanley, and F. Sciortino, 
Phys. Rev. E {\bf 69}, 050201 (2004).

\bibitem{chau}
P.L. Chau and A.J. Hardwick, Mol. Phys. {\bf 93}, 511 (1998).

\bibitem{poolepre} 
P. H. Poole, U. Essmann, F. Sciortino, and H. E. Stanley, 
Phys. Rev. E {\bf 48}, 4605 (1993). 

\bibitem{qt-LJ}
J.R. Errington, P.G. Debenedetti, and S. Torquato, 
J. Chem.  Phys. {\bf 118}, 2256 (2003).

\bibitem{mossa}
S. Mossa and F. Sciortino, Phys. Rev. Lett. {\bf 92}, 
045504 (2004).

\bibitem{franchescoTheoryPEL}
E. La Nave, S. Mossa, and F. Sciortino,
Phys. Rev. Lett. {\bf 88}, 225701 (2002).

\bibitem{kob}
W. Kob, F. Sciortino, and P. Tartaglia,
Europhys. Lett. {\bf 49}, 590 (2000).

\bibitem{Srinature}
S. Sastry, P.G. Debenedetti, and F.H. Stillinger,
Nature {\bf 393}, 554 (1998).

\bibitem{poole} 
P. H. Poole, F. Sciortino, U. Essmann and H. E. Stanley,
Nature {\bf 360}, 324 (1992). 

\bibitem{loerting} 
T. Loerting, C. Salzmann, I. Kohl, E. Mayer and A. Hallbrucker, 
Phys. Chem. Chem. Phys. {\bf 3}, 5355 (2001). 

\bibitem{finneyHDAVHDA} 
J. L. Finney, D. T. Bowron, A.K. Soper, T. Loerting, 
E. Mayer and A. Hallbrucker, 
Phys. Rev. Lett. {\bf 89}, 205503 (2002).

\bibitem{HGW}
P. Br{\"u}ggeller and E. Mayer, Nature {\bf 288}, 569 (1980).

\bibitem{bellisent}
M.-C. Bellissent-Funel, L. Bosio, A. Hallbrucker, E. Mayer, and 
R. Sridi-Dorbez, J. Chem. Phys. {\bf 97}, 1282 (1992). 

\bibitem{waterAB}
G.P. Johari, A. Hallbrucker, and Mayer, Science {\bf 273}, 90 (1996).

\bibitem{klugPRL}
D.D. Klug, C.A. Tulk, E.C. Svensson, and C.-K. Loong,
Phys. Rev. Lett. {\bf 83}, 2584 (1999).

\bibitem{TseKlug}
J.S. Tse, D.D. Klug, C.A. Tulk, I. Swainson, E.C. Svensson, C.-K. Loong,
V. Shpakov, V.R. Belosludov, R.V. Belosludov, and Y. Kawazoe,
Nature {\bf 400}, 647 (1999).

\bibitem{genePhysToday}
P.G. Debenedetti and H.E. Stanley, Phys. Today {\bf 56}, 40 (2003).







 



\end{references}
\end{document}